\documentclass[preprint,showpacs,preprintnumbers,amsmath,amssymb,floatfix,superscriptaddress]{revtex4}

\usepackage{graphicx}
\usepackage{dcolumn}
\usepackage{bm}

\newcommand{\nn}{\nonumber}
\newcommand{\ovl}[1]{\overline{#1}}

\newcommand{\p}{\partial}
\newcommand{\pslash}{p\kern-1ex /}
\newcommand{\lslash}{l\kern-1ex /}
\newcommand{\kslash}{k\kern-1ex /}
\newcommand{\dslash}{\p\kern-1.2ex /}
\newcommand{\Dslash}{{\cal D}\kern-1.5ex /}
\newcommand{\Aslash}{A\kern-1.2ex 23/}

\newcommand{\Tr}{{\rm Tr}}

\newcommand{\vev}[1]{\left\langle #1 \right\rangle}

\newcommand{\hu}[1]{{\bf }}

\newcommand{\YITP}{
  Yukawa Institute for Theoretical Physics, 
  Kyoto University, Kyoto 606-8502, Japan
}

\newcommand{\GUAS}{
  School of High Energy Accelerator Science,
  The Graduate University for Advanced Studies (Sokendai),
  Tsukuba 305-0801, Japan
}

\newcommand{\KEK}{
  High Energy Accelerator Research Organization (KEK),
  Tsukuba 305-0801, Japan
}

\newcommand{\NAGOYA}{
  Department of Physics, 
  Nagoya University, 
  Nagoya 464-8602, Japan
}

\newcommand{\NTU}{
 Physics Department, Center for Quantum Science and Engineering,
 and Center for Theoretical Sciences, National Taiwan University,
 Taipei 10617, Taiwan
}

\newcommand{\OSAKA}{
  Department of Physics, 
  Osaka University
  Toyonaka, Osaka 560-0043, Japan
}

\begin{document}

\preprint{KEK-CP-224}

\title{Non-perturbative renormalization of bilinear operators with 
dynamical overlap fermions}

\author{J.~Noaki}
\affiliation{\KEK}
\author{T.W.~Chiu}
\affiliation{\NTU}
\author{H.~Fukaya}
\affiliation{\NAGOYA}
\author{S.~Hashimoto}
\affiliation{\KEK}
\affiliation{\GUAS}
\author{H.~Matsufuru}
\affiliation{\KEK}
\author{T.~Onogi}
\altaffiliation[Present address:]{\OSAKA}
\affiliation{\YITP}
\author{E.~Shintani}
\altaffiliation[Present address:]{\OSAKA}
\affiliation{\YITP}
\author{N.~Yamada}
\affiliation{\KEK}
\affiliation{\GUAS}

\collaboration{JLQCD and TWQCD Collaboration}

\date{\today}

\begin{abstract}
Using the non-perturbative renormalization technique, we calculate the
renormalization factors for quark bilinear operators made of overlap
fermions on the lattice. 
The background gauge field is generated by the JLQCD and TWQCD
collaborations including dynamical effects of two or 2+1 flavors of
light quarks on a 16$^3\times$32 or 16$^3\times$48 lattice at lattice
spacing around 0.1~fm. 
By reducing the quark mass close to the chiral limit, where the finite
volume system enters the so-called $\epsilon$-regime, the unwanted
effect of spontaneous chiral symmetry breaking on the renormalization
factors is suppressed.
On the lattices in the conventional $p$-regime, this effect is
precisely subtracted by separately calculating the contributions from
the chiral condensate.
\end{abstract}


\maketitle

\section{Introduction}\label{introduction}
For lattice calculations of operator matrix elements including those of
electroweak effective Hamiltonian, the operator
matching is a necessary step to absorb the difference of the
renormalization scheme from the conventional continuum one, such as
the $\overline{\mathrm{MS}}$ scheme.
This is necessary for most composite operators except for those
protected by some symmetry, {\it e.g.} the conserved vector current,
since the operators are defined with a given lattice action and in
general divergent in the continuum limit.
This operator matching can be done perturbatively and has been done 
often at the 
one-loop level, which induces a potential source of large systematic
error. 
Given that the strong coupling constant $\alpha_s$ is in the range
0.2--0.3, a typical size of the two-loop correction is 4--10\%.
Non-perturbative technique to calculate this operator matching is
therefore highly desirable 
to achieve precise calculation of physical quantities. 

The Non-Perturbative 
Renormalization (NPR) method uses the RI/MOM scheme~\cite{Martinelli:1994ty}
in an intermediate step. 
This scheme is defined for the amputated Green's function in the Landau-gauge 
with an off-shell momentum, which is space-like.
Since the matching between the RI/MOM and the $\overline{\mathrm{MS}}$
schemes are known to two-loop order in many important operators, the
method provides a better matching scheme as a whole, though not the 
entire steps are non-perturbative.
Moreover, since the perturbative series is in general more convergent
in the continuum schemes, the remaining uncertainty can be made small
to a few percent level.

Since the method still requires perturbative expansion, the
renormalization condition has to be applied in the region where
non-perturbative effects are sufficiently small.
On the other hand, one has to avoid large discretization effects
that may arise when the renormalization scale is too high.
Therefore, the renormalization scale $\mu$ must satisfy the condition
$\Lambda_{QCD}\ll\mu\ll \pi/a$,
where $\Lambda_{QCD}$ stands for the QCD scale and $a$ is the lattice spacing.
This region is often called the NPR window.

The non-perturbative effect may be enhanced when the spontaneous chiral
symmetry breaking (SCSB) 
occurs and (almost) massless pions arise~\cite{Martinelli:1994ty}. 
The reason is that the pion-pole contribution in the pseudoscalar 
channel diverges towards the massless limit and makes it difficult 
to find the NPR window.
With the Wilson-type fermions, the problem is severer because the error
starts at ${\cal O}(a)$ and thus the possible window is narrower in the high
momentum regime. 
Even with the on-shell ${\cal O}(a)$-improved Wilson fermion, the problem
remains since the off-shell amplitude is considered in NPR.
With the chirally symmetric lattice actions, such as the domain-wall
and overlap fermion formulations, the problem becomes more tractable
because the ${\cal O}(a)$ error is absent even in off-shell amplitudes.

So far, there have been a number of works that calculate the
non-perturbative renormalization factors with the RI/MOM scheme for 
the domain-wall~\cite{Blum:2001sr,Aoki:2007xm} and for the quenched 
overlap fermions~\cite{DeGrand:2005af,Zhang:2005sc,Galletly:2006hq}. 

In this work, we study the non-perturbative renormalization factors 
with the RI/MOM scheme for the quark bilinear operators in unquenched QCD 
with overlap fermions.
Our motivation is two-fold. The first is to provide the
renormalization factors corresponding to the two-flavor~\cite{Aoki:2008tq} and 
2+1-flavor~\cite{Hashimoto:2007vv,Matsufuru:2008aa} gauge configurations 
generated in
the large-scale dynamical overlap project by the JLQCD and TWQCD 
collaborations, including the quark 
mass renormalization factor $Z_m$ that has been already used in a series 
of publications~\cite{Fukaya:2007yv,Fukaya:2007fb,Fukaya:2007pn,Aoki:2007pw,Chiu:2008jq,Noaki:2008iy,Noaki:2008gx}. 
The second
is to study the pion-pole contribution appearing in the NPR 
calculation in detail and demonstrate a method to control the pion-pole 
effect in a reliable manner.

Since the low-lying eigenmodes of the Dirac operator are expected to 
dominate the pion-pole contribution, it is possible to trace its effect 
as a function of
quark mass by explicitly constructing the relevant piece from the
low-mode eigenvalues.
To be explicit, the pion-pole contribution of the form
$\langle\bar{q}q\rangle/p^2$ in the operator product expansion
contains the chiral condensate $\langle\bar{q}q\rangle$, which is
finite in the vacuum of spontaneously broken chiral symmetry.
On the lattice of finite volume $V$, it quickly vanishes as quark mass 
becomes smaller than $\sim 1/\Sigma V$, where $\Sigma$ is the chiral
condensate in the infinite volume limit.
We identify this term by explicitly comparing the lattice data of the
(inverse) quark propagator with the condensate
$\langle\bar{q}q\rangle$ constructed from the eigenvalues.
Thus, this unnecessary term for NPR can be identified and subtracted.
It means that the pion-pole contribution is no longer a problem for
the NPR calculation.
Clearly, this is possible only when the chiral symmetry is preserved
on the lattice.
Otherwise, the chiral condensate has a bad cubic divergence even in
the massless limit, hence the identification of its physical
contribution is not feasible.

This paper is organized as follows.
We describe the profile of the gauge configurations used in this work
in  Section~\ref{Simulation}. 
In Section~\ref{NPRmethod}, we discuss the NPR method and its relation
to spontaneous chiral symmetry breaking and present our analysis.
Results of the calculation are given in Section~\ref{Results},
where we summarize all results of the renormalization factor
available from simple bilinear operators,
namely those for the quark mass, the scalar current, the tensor operator
and the quark field.
(The vector and axial vector currents are treated independently.)
Our conclusion is given in Section~\ref{Conclusion}.

\section{GAUGE CONFIGURATIONS}\label{Simulation}

In order to make this paper self-contained, we briefly describe the 
generation of the gauge configurations used in this work.
We refer \cite{Aoki:2008tq,Hashimoto:2007vv,Matsufuru:2008aa} for more complete
description.

We use the overlap fermion formulation
\cite{Neuberger:1997fp,Neuberger:1998wv} on the lattice for both sea
and valence quarks.
The massless overlap-Dirac operator is defined as
\begin{eqnarray}
D_{\rm ov}(0) &=& m_0
 \left(
  1+\gamma_5\cdot
  {\rm sgn}\left[H_W(-m_0)\right]
 \right),\label{overlap}
\end{eqnarray}
where $H_W(-m_0)\equiv\gamma_5 D_W(-m_0)$ is the hermitian
Wilson-Dirac operator with a large negative mass $-m_0$.
The massive operator with a bare mass $m$ is constructed from this as
\begin{eqnarray}
  D_{\rm ov}(m)
  &=& 
  \left(1-\frac{m}{2m_0}\right) D_{\rm ov}(0)
  +m.\label{Dmassive}
\end{eqnarray}
We use the Hybrid Monte Carlo (HMC) algorithm \cite{Duane:1987de}
to incorporate the fermionic determinant 
$\det [D_{\rm ov}(m_{\rm sea})]$ (for each flavor) in the path integral. 

Since the overlap-Dirac operator contains the sign function, the
corresponding determinant changes discontinuously on the border of the global
topological charge of the gauge field configuration, which makes the
simulation time-consuming.
In order to avoid touching the border, where the sign of the lowest 
eigenvalue of $H_W$ changes, we introduce two extra flavors of heavy 
Wilson fermions such that they produce a factor
\begin{eqnarray}
  \det\left[\frac{H_{\rm W}^2(-m_0)}{H_{\rm W}^2(-m_0)+\mu^2}\right]
  \label{determinant}
\end{eqnarray}
in the Boltzmann weight.
Associated (twisted-mass) bosons are also introduced with a twisted
mass $\mu$. 
They play a role to minimize the change of the effective gauge coupling
induced by those extra fermions.
Throughout this paper, we choose $m_0=1.60$ and $\mu = 0.20$
 in the lattice unit.
As a result, the topological charge $Q$ of the generated gauge
configurations is fixed to its initial value~\cite{Fukaya:2006vs}. 
In this work, we choose $Q=0$.
Although the correct sampling of the $\theta$-vacuum of QCD is spoiled due 
to the fixed topology, 
the difference is suppressed for large four-volume $V$, and it is
indeed possible to reconstruct the $\theta$-vacuum physics from 
those evaluated by the path integral in a fixed topology~\cite{Aoki:2007ka}.
In any case, such finite volume effects are irrelevant for the calculation
of the renormalization constants considered in this work, as it mainly
uses the high momentum regime.

\begin{table}
 \begin{tabular}{ccccc}
  \hline\hline
  ensemble & \makebox[3.5cm]{NF2$\epsilon$} & \makebox[3.5cm]{NF2p} 
  & \makebox[3.5cm]{NF3p-a} & \makebox[3.5cm]{NF3p-b}\\
  \hline
  $N_f$& 2 & 2 & \multicolumn{2}{c}{2+1}\\
  $\beta$       & 2.35 & 2.30 & \multicolumn{2}{c}{2.30}\\
  $a^{-1}$ [GeV]& 1.776(38) & 1.667(17) & \multicolumn{2}{c}{1.833(12)}\\
  lattice size  & $16^3\times 32$ & $16^3\times 32$ 
	  &\multicolumn{2}{c}{$16^3\times 48$}\\
   $m_{\rm sea} $ ($m_{ud}$)
  & 0.002 
  & \parbox[t]{3.3cm}{\setlength{\baselineskip}{5mm} 
    0.015, 0.025, 0.035, 0.050, 0.070, 0.100}
  & \parbox[t]{3.3cm}{\setlength{\baselineskip}{5mm} 
    0.015, 0.025, 0.035, 0.050, 0.080}
  & \parbox[t]{3.3cm}{\setlength{\baselineskip}{5mm} 
    0.015, 0.025, 0.035, 0.050, 0.100}\\
  $m_s$ & $\infty$ & $\infty$ & 0.080 & 0.100 \\
  \parbox[t]{2cm}{\setlength{\baselineskip}{5mm} 
    \ \\ $m_q$}
  & \parbox[t]{3.3cm}{\setlength{\baselineskip}{5mm} 
    0.002,\\ 0.015, 0.025, 0.035, 0.050, 0.070, 0.100}
  & \parbox[t]{3.3cm}{\setlength{\baselineskip}{5mm} 
    \ \\0.015, 0.025, 0.035, 0.050, 0.070, 0.100}
  & \parbox[t]{3.3cm}{\setlength{\baselineskip}{5mm} 
    \ \\ 0.015, 0.025, 0.035, 0.050, 0.080}
  & \parbox[t]{3.3cm}{\setlength{\baselineskip}{5mm} 
    \ \\ 0.015, 0.025, 0.035, 0.050, 0.100}\\
  \#trajectories & 2,000 & 10,000 & 2,500& 2,500\\
  \#step traj. (NPR)&
  10 & 100 &\multicolumn{2}{c}{10}\\
  \#step traj. (WTI)& see text & 20   &\multicolumn{2}{c}{5}\\
  \#low-modes  &
  $50\times 2$ & $50\times 2$  &\multicolumn{2}{c}{$80\times 2$}\\
  \# of $p_{\rm latt}$ ($(p_{\rm latt})^2$)
  &1,375 (30) &1,375 (30)&\multicolumn{2}{c}{1,875 (53)}\\
  Relevant papers 
  &\cite{Fukaya:2007yv,Fukaya:2007fb,Fukaya:2007pn}
  &\cite{Aoki:2007pw,Noaki:2008iy,Shintani:2008ga}
  &\multicolumn{2}{c}{\cite{Noaki:2008gx}}\\
  \hline\hline
 \end{tabular}
 \caption{Parameter set for each gauge ensemble NF2$\epsilon$, NF2p,
 NF3p-a and NF3p-b. Number of trajectories are common for all sea quark
 masses in each ensemble. }
\label{params}
\end{table}

In Table~\ref{params}, we list the parameter set for each gauge
ensemble on which we calculate the renormalization factors in this work.
We performed two-flavor ($N_f=2$) and 2+1-flavor ($N_f=2+1$) runs.
One of the simulations ``NF2$\epsilon$'' is in the so-called
$\epsilon$-regime of the chiral perturbation theory, which corresponds to
a very small sea quark mass so that the pion's Compton wave length is
longer than the lattice extent.
The sea quark mass $m_{\rm sea}$ = 0.002 roughly corresponds to 3~MeV
in the physical unit.
Other runs at $N_f=2$, ``NF2p'', are in the conventional
$p$-regime, where we take six values of $m_{\rm sea}$.
The 2+1-flavor runs are performed at two different values of the
strange quark mass, $m_s$ = 0.080 (``NF3p-a'') and 0.100 (``NF3p-b''), 
so that we can interpolate (or extrapolate) the data to the physical strange 
quark mass afterwards.
For each $m_s$, we take five values of sea quark mass corresponding to
the up and down quarks $m_{ud}$.

We employ the Iwasaki gauge action for the gauge part of the lattice
formulation. 
The parameter $\beta$ in the action controls the lattice spacing $a$;
we determine the value of the lattice spacing from the Sommer scale $r_0$ 
by taking $r_0$ = 0.49~fm as an input after extrapolating the lattice
data to the chiral limit $m_{\rm sea} = 0$ or $m_{ud} = 0$ at a fixed
$\beta$. 
The spatial lattice size is $16^3$ and the temporal size is 32 and 48
for the two-flavor and the 2+1-flavor runs, respectively.

The valence quark propagator on each ensemble is computed using the 
multi-shift solver at various valence quark masses.
For each ensemble in the $p$-regime, we take the same set of masses for 
the valence quark as that for the sea quarks as listed in Table~\ref{params}.
For NF2$\epsilon$, we take seven values of valence quark mass:
$m_q$ = 0.002, 0.025, 0.015, 0.035, 0.050, 0.070, and 0.100.

On each gauge configuration fixed to the Landau gauge, we compute the
quark propagator
 $S_{\rm ov}(x|x_{\rm src}) \equiv [D_{ov}(m_q)^{-1}]_{x,x_{\rm src}}$, 
where the location of the source 
$x_{\rm src}$ is typically fixed at the origin.  
To calculate the renormalization factors, we work in the
(four-dimensional) momentum space,
\begin{equation}
  S_{\rm ov}(p_{\rm latt}) 
  = \sum_{x}e^{-ip_{\rm latt}\cdot x}S_{\rm ov}(x|x_{\rm src}).
  \label{eq:qprop}
\end{equation}
To avoid possible large discretization error, we restrict the lattice
momentum $p_{\rm latt}$ such that its each element 
$p_{\rm latt}^\mu = 2\pi n_\mu/L_\mu$ does not exceed unity.
The numbers of lattice momenta satisfying this condition are listed 
in Table~\ref{params}.
Some of them are degenerate in their magnitude $(p_{\rm latt})^2$;
the number of available data point in $(p_{\rm latt})^2$ is also
listed in parentheses.
On a $16^3\times 32$ lattice, for instance, we have 1,375 different
four-momentum from the condition $-2\le n_i\le 2\ (i=1,2,3)$  
and $-5 \le n_4 \le 5$, and there are 30 different values of
$(p_{\rm latt})^2$.
When analyzing the lattice data, we first average over different
four-momenta giving an identical $(p_{\rm latt})^2$.

\section{RI/MOM renormalization on the lattice}\label{NPRmethod}

\subsection{Renormalization condition and axial-Ward-Takahashi Identity}

We consider flavor non-singlet bilinear operators of the form
$\bar{q}\Gamma q'$ with $\Gamma$ = $\gamma_\mu$, $\gamma_\mu\gamma_5$,
$I$, $\gamma_5$ and $\gamma_\mu\gamma_\nu$, that 
we call $V$, $A$, $S$, $P$ and $T$, respectively.
In the following, we may omit the prime in $q'$ that indicates that the
quark flavor is different from $q$, but the flavor non-singlet
operator is always assumed.

With the exact chiral symmetry of the overlap fermion, these operators
are multiplicatively renormalized as
\begin{equation}
  \label{eq:ren}
  (\bar{q}\Gamma q)^R(\mu) = Z_\Gamma(\mu a) (\bar{q}\Gamma q)^0,
\end{equation}
where superscripts $R$ and $0$ represent the renormalized and bare
operators, respectively.
For divergent operators $S$, $P$ and $T$, the renormalized operator
may have a dependence on the renormalization scale $\mu$.
The multiplicative renormalization factor $Z_\Gamma(\mu a)$ then
depends on the scale $\mu$, too.
For the vector and axial-vector currents, the renormalization scale
dependence is absent because of the current conservation.
In the following notation, we may drop the dependence on $\mu$,
assuming it implicitly.
The quark field $q$ is renormalized as $q^R=Z_q^{1/2}(\mu a)q^0$.

In the RI/MOM scheme~\cite{Martinelli:1994ty}, the renormalization
condition is imposed on the amputated Green's function 
$\Lambda_\Gamma(p)
=\frac{1}{12}\mathrm{Tr}[\vev{S(p)}^{-1}G_\Gamma(p)\vev{S(p)}^{-1}\Gamma]$
to satisfy
\begin{equation}
  \label{Renorm_condition}
  \Lambda_\Gamma^R(p) = Z_q^{-1}(\mu) Z_\Gamma \Lambda_\Gamma^0 (p) = 1
\end{equation}
at a space-like off-shell momentum $p^2=\mu^2$ in the chiral limit.
Here, the Green's function 
$G_\Gamma(p)=\langle q(p)| \bar{q}\Gamma q'|\bar{q}'(p)\rangle$ 
is amputated by the vacuum expectation value of the quark propagator 
$\vev{S(p)}$ and projected with an
appropriate gamma matrix $\Gamma$.
(The `Tr' denotes the trace over the color and spinor indices.)
The RI/MOM scheme is defined for the momentum configuration that the
in-coming and out-going quark momenta are the same $p$.
Since the definition involves the external quark field, which is not
gauge invariant, the renormalization condition depends on the gauge.
In the RI/MOM scheme, the Landau gauge is chosen. 

In the RI/MOM scheme, the wave function renormalization $Z_q$ is fixed
by imposing the condition 
\begin{equation}
  \frac{1}{12i}\mathrm{Tr}
  \left[
    \frac{\partial \vev{S^R(p)}^{-1}}{\partial \pslash}
  \right]
  =
  Z_q^{-1} \frac{1}{12i}\mathrm{Tr}
  \left[
    \frac{\partial \vev{S(p)}^{-1}}{\partial \pslash}
  \right]
  = 1
\end{equation}
at $p^2=\mu^2$ in the chiral limit.
Numerically, though, this is not straightforward since it involves a
numerical derivative in terms of $p_\mu$.
Instead, we obtain $Z_q$ using (\ref{Renorm_condition}) for the
axial-vector vertex function $\Lambda_A(p)$ with an input of $Z_A$
obtained through the axial-Ward-Takahashi identity
\begin{equation}
  Z_A^{\rm WTI} \Delta_4 \langle A_4(x) {\cal O}(0)\rangle = 
  2 m_q \langle P(x) {\cal O}(0)\rangle,\label{AWTI}
\end{equation}
where $A_4$ and $P$ are the axial-vector current
in the time direction and pseudo-scalar density, respectively.
$\Delta_4$ denotes the symmetrized difference.
This relation must be satisfied as far as the position $x$ of the
operator is not too close to the origin, where some interpolating
field ${\cal O}$ is set.
Once $Z_A$ is fixed from this relation, the wave function
renormalization is determined as
$Z_q^R(\mu) = Z_A^{\rm WTI}\Lambda_A(p)$
at $p^2=\mu^2$.

\begin{table}[tbp]
 \begin{tabular}{cccccc}
  \hline\hline
  \multicolumn{2}{c}{NF2p}
 &\multicolumn{2}{c}{NF3p-a}   
 &\multicolumn{2}{c}{NF3p-b}\\
 \hline
 $m_{\rm sea}$& $Z_A^{\rm WTI}$\ \ \
& $m_{ud}$     & $Z_A^{\rm WTI}$\ \ \
& $m_{ud}$     & $Z_A^{\rm WTI}$\\
 0.015& 1.37867(61)\ \ \ & 0.015& 1.38934(49) \ \ \ & 0.015& 1.38968(47)\\
 0.025& 1.37703(45)\ \ \ & 0.025& 1.38709(40) \ \ \ & 0.025& 1.38700(36)\\
 0.035& 1.37412(40)\ \ \ & 0.035& 1.38431(32) \ \ \ & 0.035& 1.38408(32)\\
 0.050& 1.37032(33)\ \ \ & 0.050& 1.38031(27) \ \ \ & 0.050& 1.38019(31)\\
 0.070& 1.36441(31)\ \ \ & 0.080& 1.37196(21) \ \ \ & 0.100& 1.36658(26)\\
 0.100& 1.35436(29)\ \ \ &      &                   &      &            \\
 \hline
 0.00  & 1.38222(82)\ \ \  
 &\multicolumn{4}{c}{``chiral limit'': 1.39360(48)}\\
  ($\chi^2/$dof = & 0.43 & &\multicolumn{2}{c}{0.16)}\\
 \hline\hline
 \end{tabular}
 \caption{Summary of the results of $Z_A^{\rm WTI}$ as a function of 
 $m_{\rm sea}$ or $m_{ud}$ for NF2p, NF3p-a and NF3p-b. 
 The last two rows show the results of the extrapolation to the chiral 
 limit as described in the text.}
 \label{ZAsummary}
\end{table}

\begin{figure}[tbp]
 \includegraphics[width=8cm,clip]{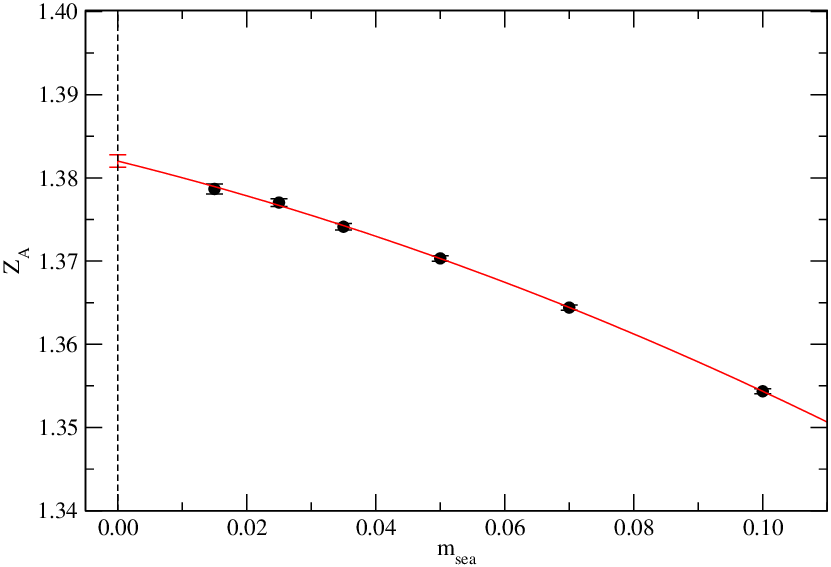}
 \includegraphics[width=8cm,clip]{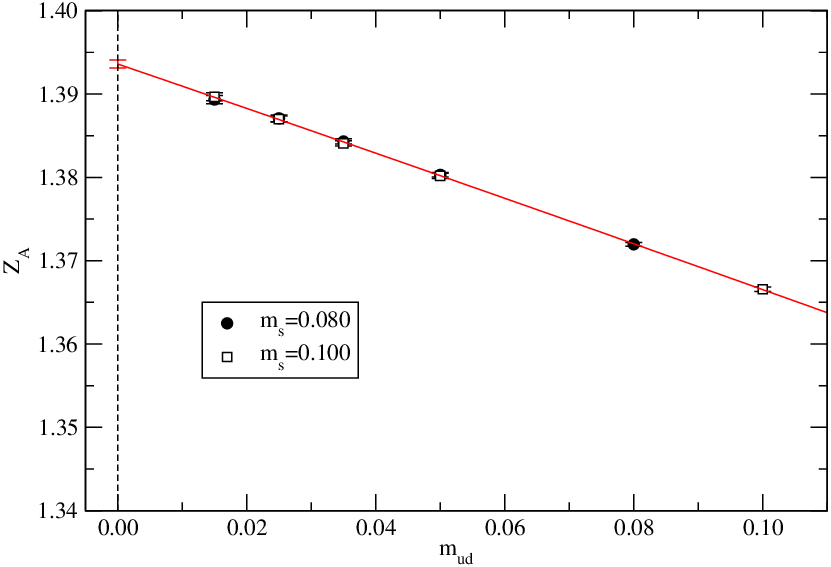}
 \caption{Chiral extrapolation of $Z_A^{\rm WTI}$ for NF2p (left panel)
 and NF3p-a and NF3p-b (right panel).}
 \label{ZAfit}
\end{figure}

In practice, we use a pseudo-scalar density with a smeared operator
for ${\cal O}$ and sum over spatial lattice sites.
Then, we fit a ratio
$2m_q \sum_{\vec{x}}\langle P(\vec{x},t){\cal O}(0)\rangle /
\Delta_t \sum_{\vec{x}}\langle A_4(\vec{x},t){\cal O}(0)\rangle$
with time slices $t \ge t_0$, which is large enough to obtain 
a constant $Z_A^{\rm WTI}$.
For NF2$\epsilon$, setting $m_q = m_{\rm sea} =0.002$ and
$t_0=4$, we obtain 
\begin{eqnarray}
 Z_A^{\rm WTI} = 1.3511(12).
\end{eqnarray}
In other ensembles, $Z_A^{\rm WTI}$ is obtained for each sea quark mass with 
the valence quark mass equal to the sea (up and down) quark mass.  
Results with $t_0=7$ for all ensembles are summarized in Table~\ref{ZAsummary}, 
where the second row from the last lists the values extrapolated 
to the chiral limit.
In the chiral extrapolation, we assume linear plus quadratic dependence
on $m_q$. 
Since the local axial-vector current we use on the lattice is not a 
conserved current at finite lattice spacings, the Ward-Takahashi identity
(\ref{AWTI}) may be slightly violated. To be explicit, a discretization effect
of the form $a^2 m_q\partial_\mu P$ is possible as an additive correction to
$A_\mu$, which leads to the linear dependence on $m_q$.
Including possible quadratic quark mass dependence, we use 
\begin{eqnarray}
 Z_A^{\rm WTI}(m_q) 
  &=& Z_A^{\rm WTI}(0)+ C_1 m_q + C_2 m_q^2.
\end{eqnarray}
by setting the valence quark mass as $m_q=m_{\rm sea}$ for NF2p and 
as $m_q=m_{ud}$ for the combined data of NF3p-a and NF3p-b.
For the case of $N_f=2+1$, we assume independence of $Z_A$ on $m_s$,
which appears only as a sea quark. This assumption is indeed supported 
by the lattice data at two different $m_s$.

The vertex function $\Lambda_\Gamma(p)$ is calculated on the lattice 
at many different momentum values $p_{\rm latt}$, whose number is
listed in Table~\ref{params}.
With the overlap fermion, we compute the vertex functions as
\begin{equation}
 \Lambda_\Gamma (p_{\rm latt}) =
 \frac{1}{12}
 \Tr\left[ 
   \langle\hat{S}_{\rm ov}(p_{\rm latt})\rangle^{-1}
   \left\langle 
     \hat{S}_{\rm ov}(p_{\rm latt})\Gamma
     \gamma_5 \hat{S}_{\rm ov}^\dagger(p_{\rm latt})\gamma_5
   \right\rangle
   \langle \gamma_5 \hat{S}_{\rm ov}^\dagger(p_{\rm latt}) \gamma_5 \rangle^{-1}
   \Gamma 
 \right]
 \label{eq:vertex}
\end{equation}
where the quark propagator is effectively given as 
\begin{equation}
  \hat{S}_{\rm ov}(p) = \frac{2m_0}{2m_0 -m_q}
  \left(
    S_{\rm ov}(p)-\frac{e^{-ip\cdot x_{\rm src}}}{2m_0}
  \right). 
 \label{Rotation}
\end{equation}
This modification of the quark propagator from $S_{\rm ov}(p)$, 
the inverse of the overlap operator $D_{\rm ov}(m_q)$, is made in order to
incorporate the quark field rotation
$q \to (1-\tfrac{D_{\rm ov}(0)}{2m_0})q$, $\bar{q}\to\bar{q}$,
which is necessary to remove the ${\cal O}(a)$ effects from off-shell
quantities. 
In (\ref{eq:vertex}), we note that 
$\gamma_5 \hat{S}_{\rm ov}^\dagger(p_{\rm latt})\gamma_5$
cannot be simply replaced by
$\hat{S}_{\rm ov}(p_{\rm latt})$
since the l.h.s of (\ref{eq:qprop}) still depends on the source point 
$x_{\rm src}$.

\subsection{Vector and axial-vector vertex functions}

\begin{figure}[tbp]
 \includegraphics[width=8cm,clip]{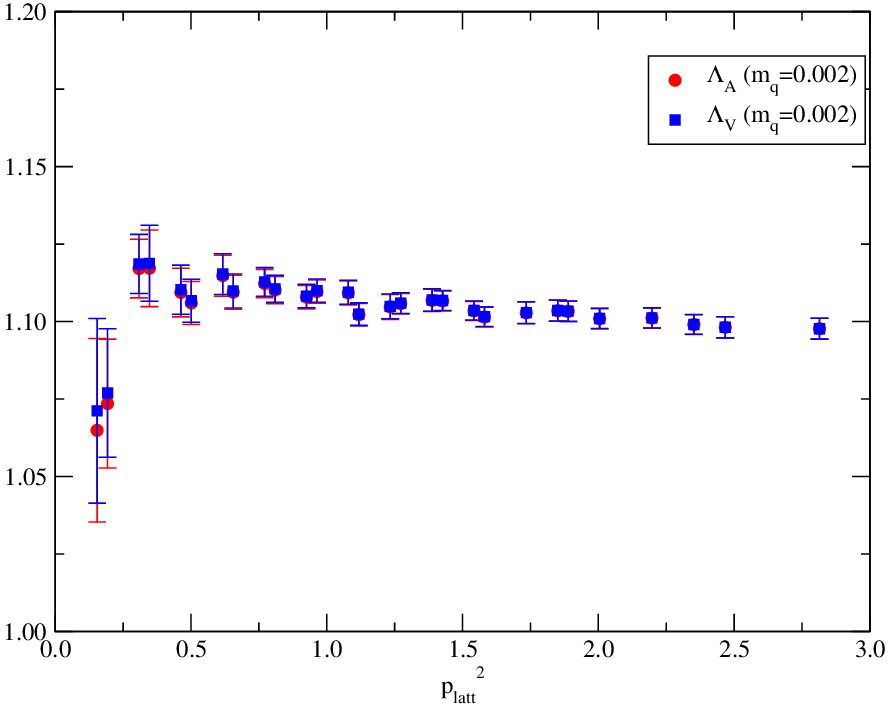}
 \includegraphics[width=8cm,clip]{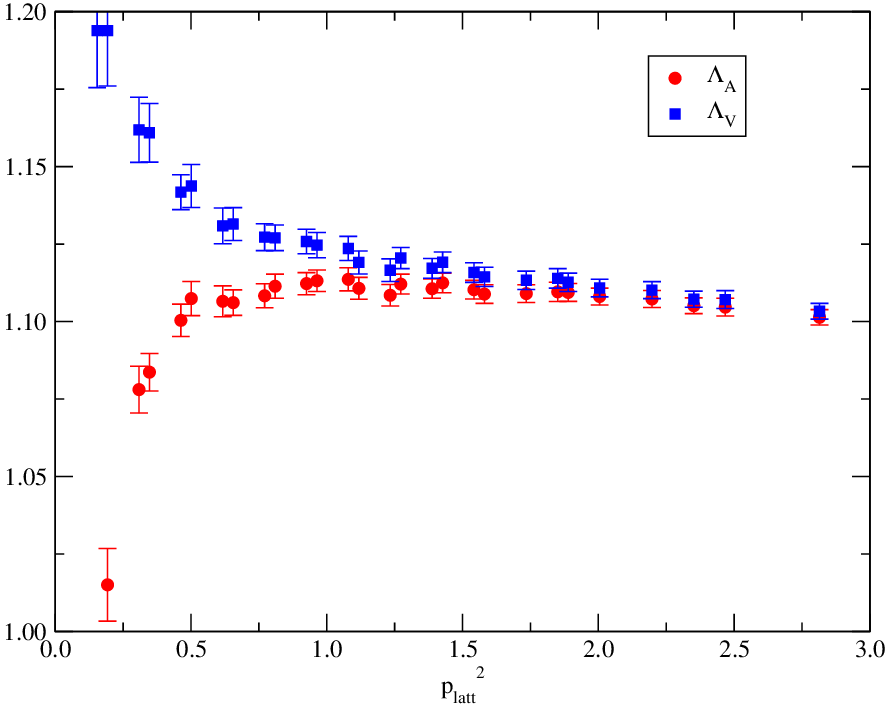}
 \includegraphics[width=8cm,clip]{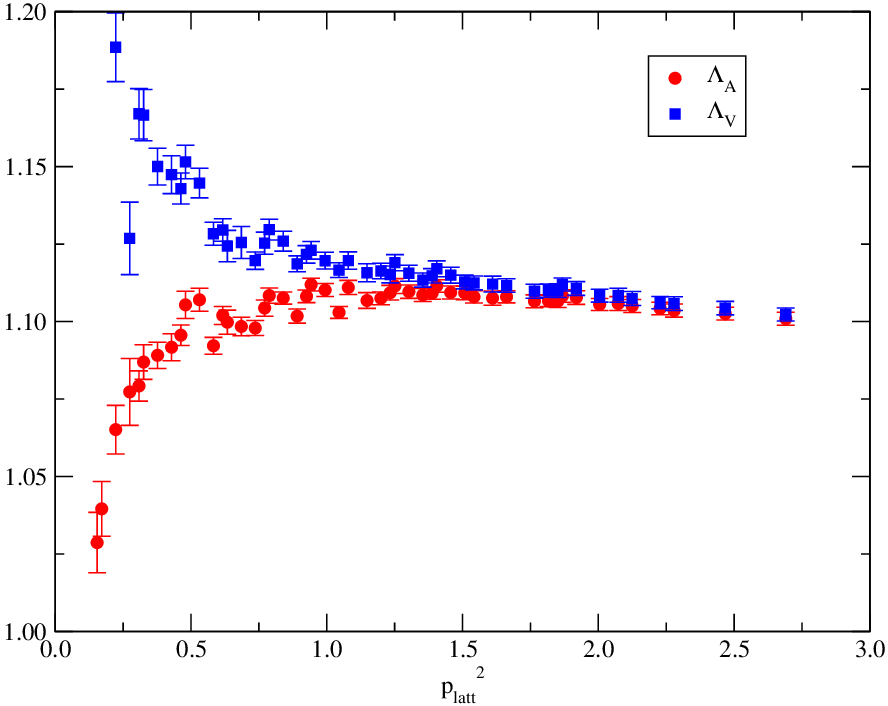}
 \includegraphics[width=8cm,clip]{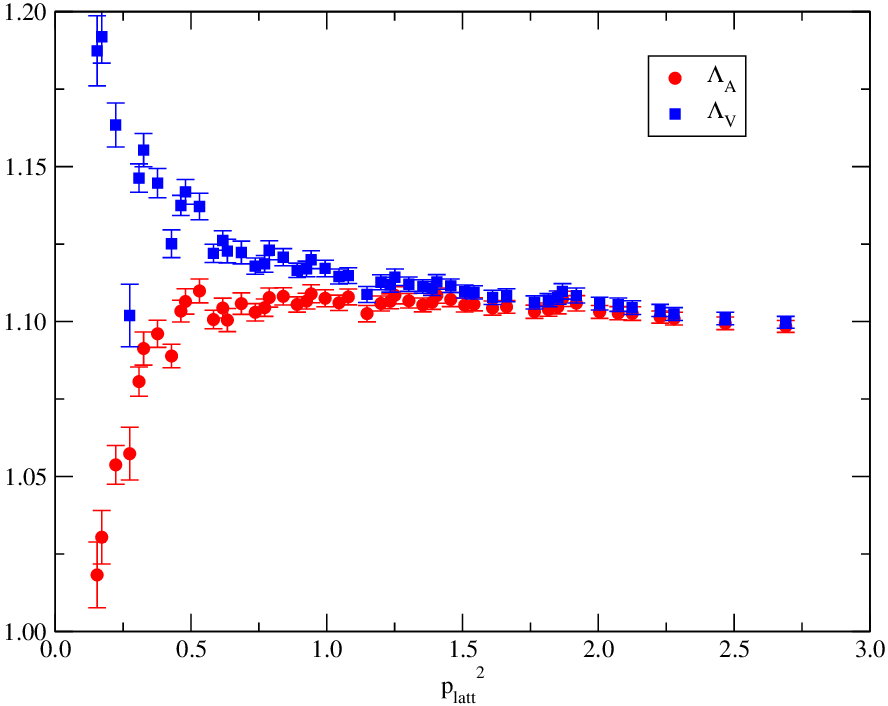}
 \caption{
   Vertex functions $\Lambda^{\rm latt}_A(p_{\rm latt})$ (circles) and 
   $\Lambda^{\rm latt}_V(p_{\rm latt})$ (squares) 
   at $m_q=0.002$ for NF2$\epsilon$ (upper left) and those 
   in the limit of massless valence quark ($m_q=0$) with the sea quark 
   mass fixed at $m_{\rm sea} (m_{ud}) = 0.015$ for NF2p (upper right), 
   NF3p-a (lower left) and NF3p-b (lower right). 
 }
 \label{VAvp}
\end{figure}

Results for the vector and axial-vector vertex functions are shown in
Figure~\ref{VAvp} as a function of $p_{\rm latt}^2 \equiv(\mu a)^2$. 
In the figure, panels from NF2p, NF3p-a and NF3p-b show the data from the 
lightest $m_{\rm sea}$ or $m_{ud}$.
The chiral symmetry implies that these two functions become identical
in the massless limit unless the symmetry is spontaneously broken.
With exact chiral symmetry of the overlap fermion, this should be 
the case even at finite lattice spacings.
The result in the $\epsilon$-regime (NF2$\epsilon$, upper-left in the
figure) clearly shows this behavior, which is consistent with the
absence of spontaneous symmetry breaking on a finite volume lattice.

Other three panels, that are obtained in the $p$-regime,
show the splitting between the vector and the axial-vector channels. 
The numerical data in these plots are naively extrapolated to the
chiral limit of the valence quarks by assuming a linear dependence on
$m_q$, but the qualitative picture remains unchanged for each
valence quark mass.

This inconsistency among the vector and axial-vector currents may be 
explained as an effect of the spontaneously broken chiral symmetry.
Even on a finite volume lattice, the spontaneous symmetry breaking
induces non-zero value of the chiral condensate
$-\langle\bar{q}q\rangle\equiv\Sigma$ as far as the quark mass
is much larger than a typical scale $1/\Sigma V$.
An Operator Product Expansion (OPE) analysis \cite{Aoki:2007xm}
suggests that there are contributions of the form 
$\Lambda_{\rm QCD}^2/p^2$ and $m\Lambda_{\rm QCD}/p^2$ to the
difference $\Lambda_A(p)-\Lambda_V(p)$. 
These contributions are induced when the momentum assignment for the
three-point function gives vanishing momentum transfer at the vertex.
Namely, when the incoming and outgoing momenta are identical as in the
RI/MOM-scheme momentum set-up, which is called the ``exceptional momenta'',
the higher dimensional terms in OPE like
$(\bar{q}q)^2/p^6$ 
(with some gamma matrices inserted in the numerator) 
may lead to a much larger contribution of the form
$\langle(\bar{q}q)^2\rangle/(\Lambda_{\rm QCD}^4p^2)$, 
which remains in the chiral limit in contrast to the lower order
contributions $m^2/p^2$ or $m\langle\bar{q}q\rangle/p^4$
\cite{Aoki:2007xm}.
This problem can be avoided by choosing other momentum configurations,
such as the RI/SMOM scheme considered in \cite{Sturm:2009kb}.

We do not go into details of this problem.
But, since the effect becomes statistically significant only below
$p_{\rm latt}^2\sim$ 1.0--1.5 for the vector and axial-vector
channels, we simply use the region that is not largely affected by
this effect in the following analysis.

\begin{figure}[tbp]
 \includegraphics[width=8cm,clip]{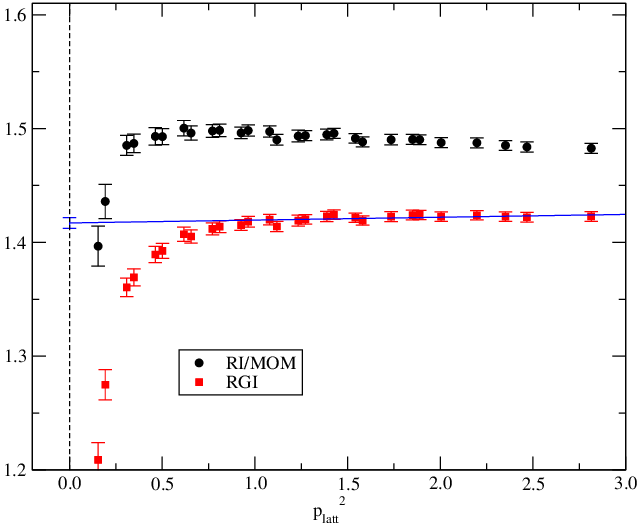}
 \includegraphics[width=8cm,clip]{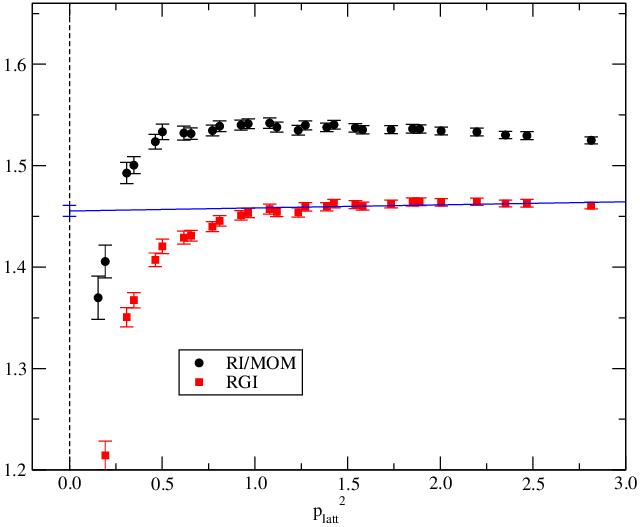}
 \includegraphics[width=8cm,clip]{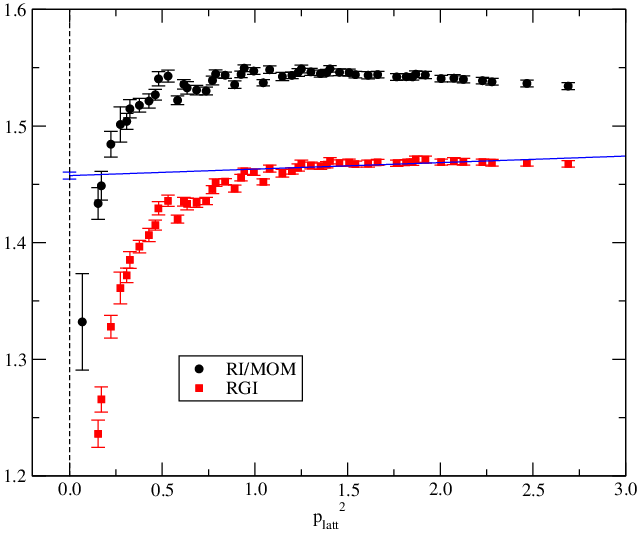}
 \includegraphics[width=8cm,clip]{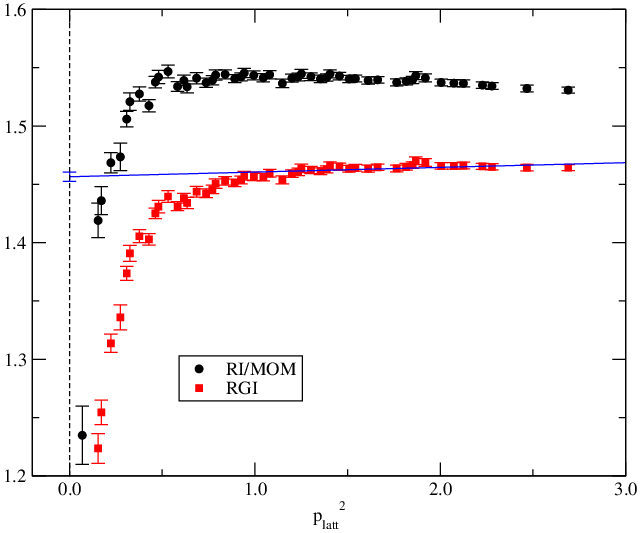}
 \caption{
   Quark field renormalization factor $Z_q$ as a function of 
   $p_{\rm latt}^2$.
   For NF2p, NF3p-a and NF3p-b, data at $m_{\rm sea} (m_{ud})$ = 0.015 are
   plotted as an example.
   Results in the RI/MOM scheme are shown by circles, while those in
   RGI are plotted by squares.
 } 
 \label{Matching_q}
\end{figure}

\begin{figure}[tbp]
 \includegraphics[width=8cm,clip]{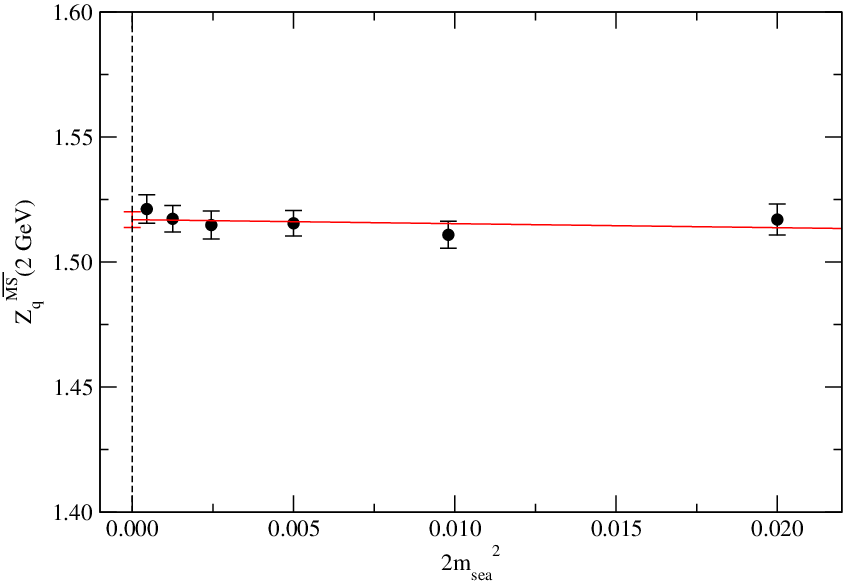}
 \includegraphics[width=8.2cm,clip]{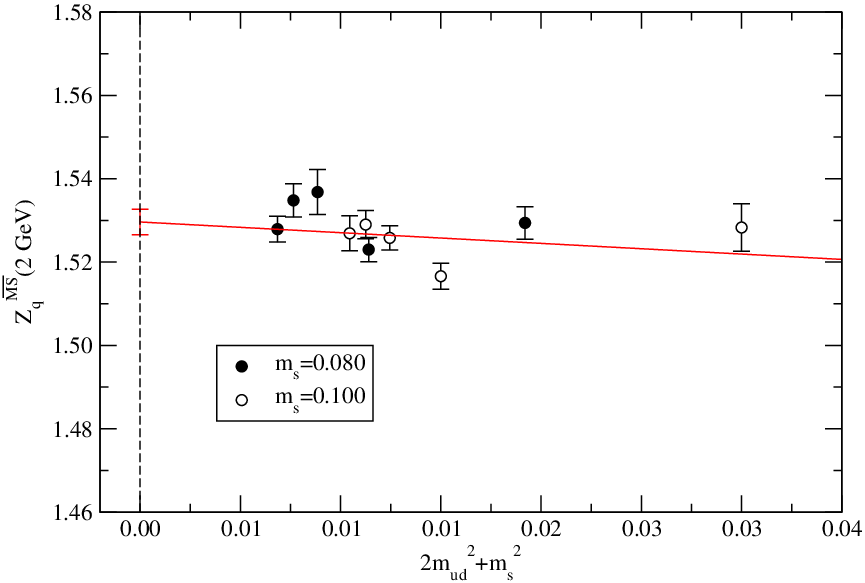}
 \caption{
   Chiral extrapolation of $Z_q^{\ovl{\rm MS}}(2\ {\rm GeV})$
   on the NF2p (left) and NF3p-a (right, filled symbols) and NF3p-b
 (right, open symbols) lattices.
}
 \label{Zqmsb_chi}
\end{figure}

\begin{table}
 \begin{tabular}{rcc}
  \hline\hline
  $m_{\rm sea}(m_{ud})$
  &\ \ \ \ $Z_q^{\rm RGI}$\ \ \ \ 
  & $Z_q^{\ovl{\rm MS}}(2\ {\rm GeV})$ \\
  \hline
  NF2$\epsilon$,\  
  0.002\ \ &1.4170(47) &1.4799(50) \\
  \hline\hline
  NF2p,\ \
  0.015\ \ & 1.4540(54)&  1.5186(56)\\
  0.025\ \ & 1.4503(51)&  1.5147(53)\\
  0.035\ \ & 1.4479(53)&  1.5122(56)\\
  0.050\ \ & 1.4486(49)&  1.5129(51)\\
  0.070\ \ & 1.4442(51)&  1.5083(54)\\
  0.100\ \ & 1.4500(59)&  1.5143(62)\\
  \hline
  0.000\ \ & 1.4526(30)&  1.5170(31)\\
  ($\chi^2/$dof&  0.33      &  0.33 ) \\
  \hline\hline
  NF3p-a,\ \
  0.015\ \ &1.4575(30)& 1.5279(31)\\
  0.025\ \ &1.4641(38)& 1.5348(40)\\
  0.035\ \ &1.4660(51)& 1.5368(54)\\
  0.050\ \ &1.4528(27)& 1.5230(29)\\
  0.080\ \ &1.4590(37)& 1.5294(39)\\
  \hline
  NF3p-b,\ \
  0.015\ \ &1.4565(40)& 1.5269(42)\\
  0.025\ \ &1.4585(32)& 1.5290(34)\\
  0.035\ \ &1.4555(28)& 1.5258(29)\\
  0.050\ \ &1.4467(30)& 1.5166(31)\\
  0.100\ \ &1.4578(55)& 1.5283(57)\\
  \hline
  ``chiral limit'':\ \ & 1.4592(29) &1.5296(31)\\
  ($\chi^2/$dof &  2.20       &  2.20  ) \\
  \hline\hline
 \end{tabular}
 \caption{
   Numerical results for the quark wave function renormalization
   factor $Z_q$.
   The values in the RGI definition $Z_q^{\rm RGI}$ and those defined
   in the $\overline{\rm MS}$ scheme at $\mu$ = 2~GeV,
   $Z_q^{\overline{\rm MS}}(2{\rm ~GeV})$ are listed for each sea
   quark mass.
 }
 \label{Zfactors_q}
\end{table}

The quark field renormalization factor $Z_q(\mu)$ can be obtained from 
$\Lambda_A(p)$ by multiplying the axial-current renormalization
constant $Z_A^{\rm WTI}$ as determined from the Ward-Takahashi
identity.
The results are shown in Figure~\ref{Matching_q} by filled circles
as a function of $p_{\rm latt}^2$.
The different panels represent the data from the ensembles
NF2$\epsilon$, NF2p, NF3p-a, and NF3p-b, respectively.
By multiplying the matching factor $1/w_q^{\rm RI/MOM}(q)$ at the 
four-loop level as defined
in Appendix~\ref{matching}, we may define the Renormalization Group
Invariant (RGI) quantity, which is also scheme independent.
Our numerical results plotted by squares in Figure~\ref{Matching_q}
clearly show the expected scale independence.
Since we expect discretization effects proportional to 
$a^2p_{\rm latt}^2$, we fit the lattice data above $(p_{\rm latt})^2$
= 1.0 by a linear function and obtain the result for $Z_q^{\rm RGI}$
from an intercept at $p_{\rm latt}^2=0$.
The lattice data below $a^2p_{\rm latt}^2\simeq$ 1.0 are largely
affected by the effect of spontaneously broken chiral symmetry and
deviate from the linear behavior as expected.

The results for $Z_q^{\rm RGI}$ are listed in Table~\ref{Zfactors_q}.
Also listed are the results converted to the $\overline{\rm MS}$
scheme at $\mu$ = 2~GeV using the four-loop level matching constant 
$w_q^{\overline{\rm MS}}(\mu)$ defined in Appendix~\ref{matching}. 

So far, the results are given at each sea quark mass after taking
the chiral limit of valence quarks.
The chiral limit of sea quarks can be taken by assuming that the
sea quark mass dependence has the form 
$Z(1+2c^{(2)} m_{\rm sea}^2)$ (for NF2p) or 
$Z(1+c^{(3)} (2m_{ud}^2+m_s^2))$ (for NF3p-a and NF3p-b).
The coefficients $c^{(2)}$ and $c^{(3)}$ are numerical constants
depending on the number of flavors.
The linear term in $m_{\rm sea}$ (or in $m_{ud}$) should not remain
for the quantities irrelevant to the chiral symmetry breaking.
Figure~\ref{Zqmsb_chi} shows the chiral extrapolation of
$Z_q^{\overline{\rm MS}}(2\mathrm{~GeV})$ for both NF2p and NF3p-a/NF3p-b.
We do not observe any significant sea quark mass dependence.
The chiral extrapolation should therefore be very stable.
The results are listed in Table~\ref{Zfactors_q}.

\subsection{Scalar and Pseudo-scalar vertex functions}\label{PBPtreatments}

\begin{figure}
 \includegraphics[width=8cm,clip]{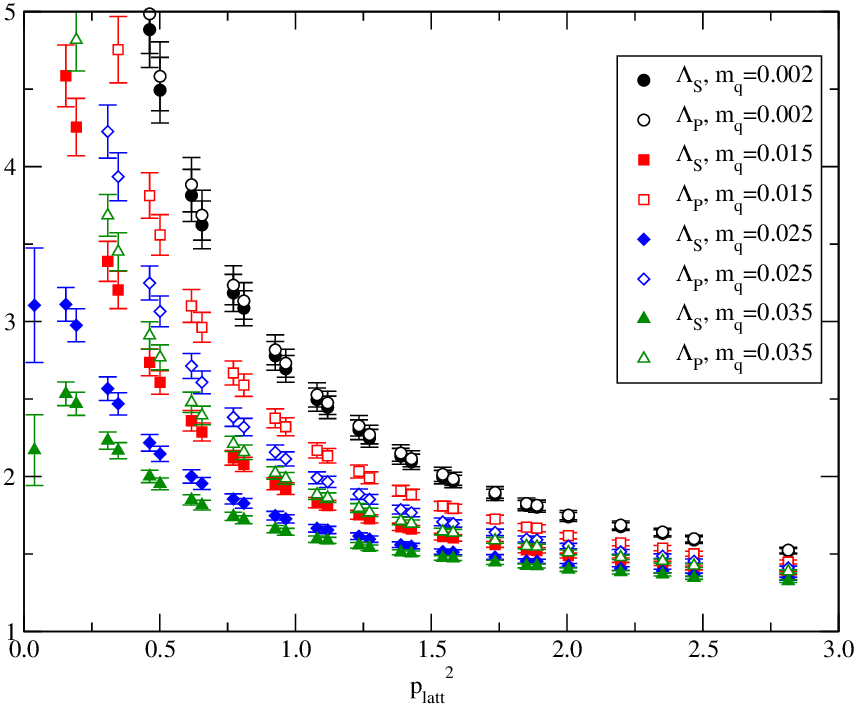}
 \includegraphics[width=8cm,clip]{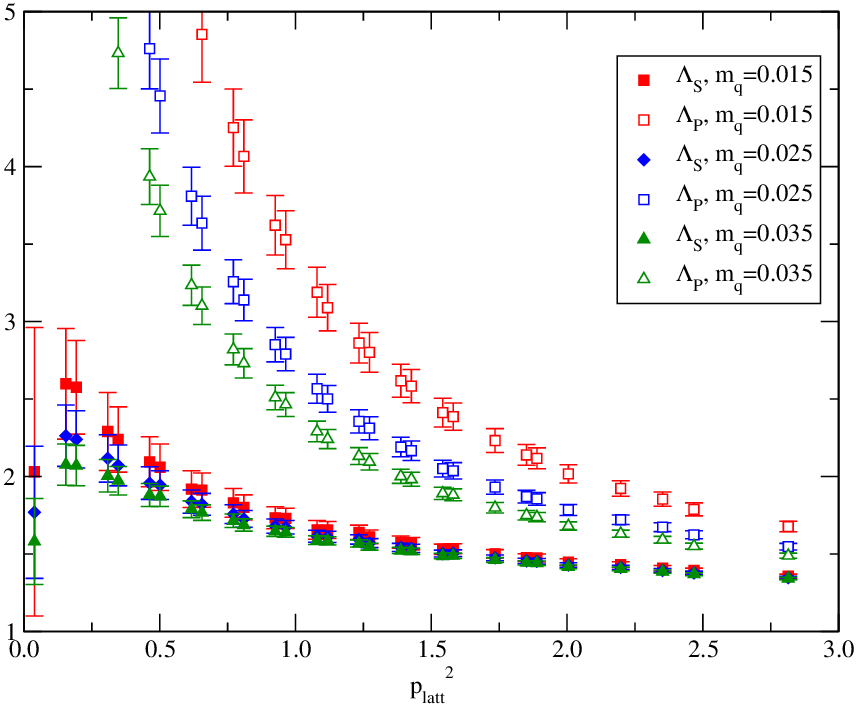}
 \includegraphics[width=8cm,clip]{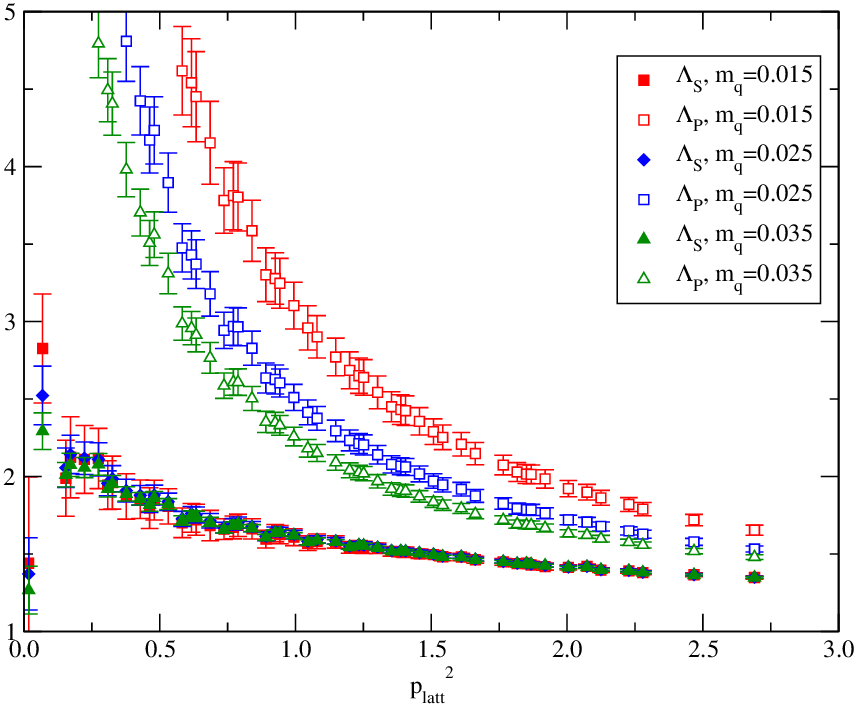}
 \includegraphics[width=8cm,clip]{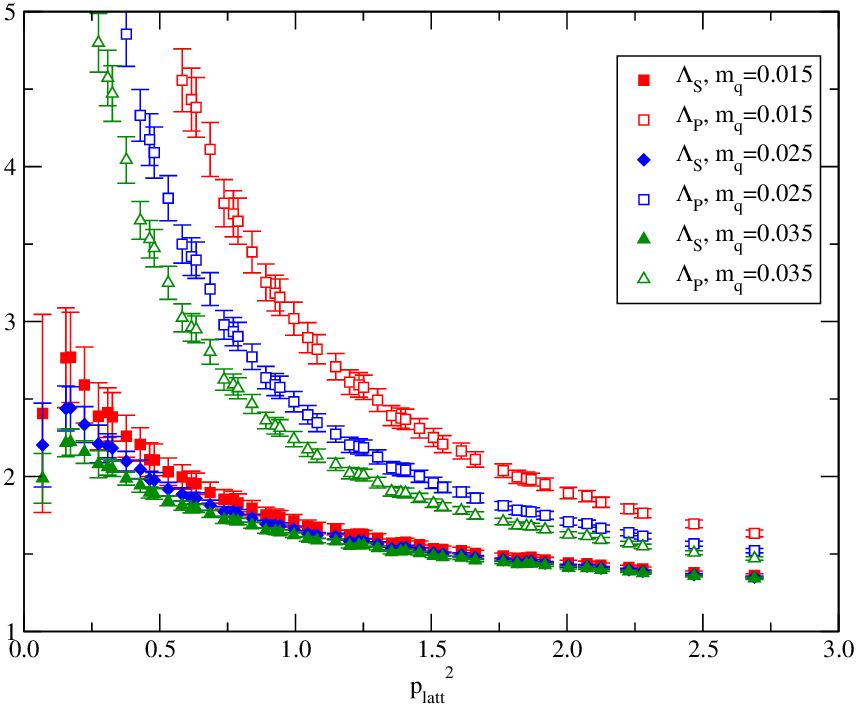}
 \caption{
   Vertex functions $\Lambda^{\rm latt}_S(p_{\rm latt})$ 
   (filled symbols) and $\Lambda^{\rm latt}_P(p_{\rm latt})$ (open
   symbols).
   The data are shown for NF2$\epsilon$ (upper left), NF2p (upper
   right), NF3p-a (lower left) and NF3p-b (lower right) as functions of
   $p_{\rm latt}^2$. 
   In each panel, data at $m_q=0.015$, $0.025$ and 
   $0.035$ are shown by squares, diamonds and triangles,
   respectively. 
   For NF2$\epsilon$, data at $m_q=0.002$ are shown as well by
   circles. 
   For NF2p, NF3p-a and NF3p-b, data at the lightest sea quark 
   $m_{\rm sea} (m_{ud}) = 0.015$ are plotted as an example.
 }
 \label{SPvp}
\end{figure}

In Figure~\ref{SPvp}, the momentum dependence of the scalar vertex
function $\Lambda_S^{\rm latt}(p_{\rm latt})$ (filled symbols) and 
the pseudo-scalar vertex function
$\Lambda_P^{\rm latt}(p_{\rm latt})$ (open symbols) is shown for 
each ensemble. 
For the data in the $\epsilon$-regime (circles in the upper left
panel), we observe an excellent agreement between $\Lambda_S(p)$ (filled
symbols) and $\Lambda_P(p)$ (open symbols), which is expected from the
exact chiral symmetry of the overlap fermion.
On the other hand, once the valence quark mass $m_q$ is out of the
$\epsilon$-regime (the data at $m_q$ = 0.015, 0.025 and 0.035 are
plotted by squares, diamonds and triangles, respectively), we find
large disagreement between $\Lambda_S(p)$ and $\Lambda_P(p)$.

This observation again indicates the effect of the spontaneous chiral
symmetry breaking.
From the OPE analysis one expects that this effect is more
significant than in $\Lambda_V(p)$ and $\Lambda_A(p)$, because the
violation is enhanced by an inverse quark mass as discussed below.
One has to subtract this effect in order to extract the
renormalization constants because its matching is based on the continuum
perturbation theory that does not contain non-perturbative effects.

We consider the quark mass dependence of 
$\Lambda_S(p)$ and $\Lambda_P(p)$ using OPE along the line of the
analysis in \cite{Blum:2001sr}.
Using the vector and axial-vector Ward-Takahashi identities,
one may obtain relations between the vertex functions and the inverse
quark propagators as \cite{Giusti:2000jr}
\begin{eqnarray}
  \label{eq:WTI_S}
 \Lambda_S (p) &=& 
  \frac{1}{12}\frac{\partial\, \Tr \vev{S(p)}^{-1}}{\partial m_q},\\
  \label{eq:WTI_P}
 \Lambda_P (p) &=& 
  \frac{1}{12}\frac{\Tr \vev{S(p)}^{-1}}{m_q}.
\end{eqnarray}
On the lattice we use the improved overlap quark propagator 
$\hat{S}_{\rm ov}(p)$ in place of $S(p)$.
From OPE the inverse quark propagator $\Tr\vev{S(p)}^{-1}$ may be written
as \cite{Politzer:1976tv} 
\begin{equation}
  \frac{1}{12}\Tr \vev{S(p)}^{-1}
   = C\cdot\frac{\vev{\bar{q}q}}{p^2} +Z_qZ_m m_q +\cdots
\end{equation}
in the large $p^2$ regime.
The effect of the chiral symmetry breaking is picked up through the
chiral condensate $\vev{\bar{q}q}$, and 
$C$ is a perturbatively calculable constant.
At the one-loop level, $C=4\pi\alpha_s/3$.
As in the case of the vector and axial-vector vertex functions, the
effects from higher dimensional operators, such as
$\langle(\bar{q}q)^2\rangle$, may also exist.
They are usually suppressed by additional powers of $1/p^2$, but
due to the lack of the momentum injection the suppression may not
work in the case of the inverse quark propagator and of the vertex
functions at zero momentum transfer.
We therefore leave $C$ as an unknown constant instead of using the
perturbatively known value.

Then, using the relations (\ref{eq:WTI_S}) and (\ref{eq:WTI_P}), we
may evaluate the effect of the chiral condensate on the vertex
functions as 
\begin{eqnarray}
 \Lambda_S(p) 
  &=& \frac{C}{p^2} \frac{\partial \vev{\bar{q}q}}{\partial m_q}
  +Z_qZ_m + \cdots,
  \label{LmdS}\\
 \Lambda_P(p) 
  &=& \frac{C}{p^2} \frac{\vev{\bar{q}q}}{m_q} +Z_qZ_m + \cdots.
 \label{LmdP}
\end{eqnarray}
From these expressions, one sees an enhancement in the low $p^2$
region due to the chiral condensate only for the pseudo-scalar
channel, while the scalar channel should not be affected too much
because of a derivative with respect to $m_q$ rather than a factor
$1/m_q$. 

The quark mass dependence of the condensate $\vev{\bar{q}q}$ is not a
trivial issue, since it has effects from both ultraviolet and infrared
origins.
Since the operator $\bar{q}q$ contains quadratic divergence of the form
$m_q/a^2$ apart from the chiral limit, the chiral condensate
$\vev{\bar{q}q}$ directly calculated on the lattice contains
unphysical large $m_q$ dependence.
It has to be subtracted before the analysis using (\ref{LmdS}) and
(\ref{LmdP}), because the formulae are obtained as an expansion around
the chiral limit.

In the infrared regime, the chiral condensate has a non-trivial
quark mass dependence especially in a finite volume.
First, because of the pion-loop effects, the chiral condensate
develops the chiral logarithm of the form $m_q\ln m_q$ with known 
coefficients~\cite{Gasser:1983yg}.
On a finite volume lattice, the quark mass dependence becomes more 
complicated.
Namely, once the quark mass enters the $\epsilon$-regime, the mass
dependence is no longer governed by the simple chiral logarithm, but
given by the formula recently developed in~\cite{Damgaard:2008zs}.

In our analysis, instead of using the formula in~\cite{Damgaard:2008zs}
we calculate the condensate using its eigenvalue decomposition
by making use of the low eigenmodes obtained on the same ensembles.
For each lattice configuration, we define
\begin{eqnarray}
  \label{eq:condensate}
  (\bar{q}q)^{(N)}= \frac{1}{L_s^3L_t}\sum_{i=1}^{N}
  \frac{2m_q}{m_q^2 +\hat{\lambda}_i^*\hat{\lambda}_i},
\end{eqnarray}
where $\hat{\lambda}_i$ is an eigenvalue of the massless overlap-Dirac
operator, which satisfies the eigen equation
\begin{equation}
  D_{\rm ov}(0) \left(1-\frac{D_{\rm ov}(0)}{2m_0}\right)^{-1} u_i(x) = 
  \hat{\lambda}_i u_i(x)
\end{equation}
with $u_i(x)$ an eigenvector.
In (\ref{eq:condensate}) we use the fact that the eigenvalues appear
as complex conjugate pairs. 
The normalization in (\ref{eq:condensate}) contains the lattice volume
$L_s^3 L_t$.

We truncate the sum in (\ref{eq:condensate}) at $N$-th eigenvalue,
which may be considered as a ``renormalization scheme'' to define the
divergent operator $\bar{q}q$.
Here, $N$ plays a role of the ultraviolet cut-off.
After taking an ensemble average, we denote the chiral condensate thus
defined as $\langle\bar{q}q\rangle^{(N)}$.
In the course of our project, we calculate and store the low-lying
eigenvalues and eigenvectors of the overlap-Dirac operator. 
In addition to the calculation of the truncated chiral condensate 
(\ref{eq:condensate}),
these eigenmodes can be used to precondition the solvers, to average over 
source points, or to construct disconnected diagrams in 
the calculations of physical observables
~\cite{Aoki:2007pw,Noaki:2008iy,JLQCD:2009qn}.
The numbers of the stored low-modes for each configuration are listed
in Table~\ref{params}.

From a dimensional analysis, the quark mass dependence of
$\langle\bar{q}q\rangle^{(N)}$ may be parametrized as 
\begin{equation}
  \vev{\bar{q}q}^{(N)} = 
  \vev{\bar{q}q}^{\rm (subt)} +  c_1^{(N)} \frac{m_q}{a^2} + c_2^{(N)} m_q^3.
  \label{pbpN}
\end{equation}
Because of the exact chiral symmetry of the overlap fermion, there is
no leading power divergence of the order $1/a^3$, and the term behaves as
$m_q^2/a$ is also absent. 
Although the cubic term $c_2^{(N)}m_q^3$ in (\ref{pbpN}) may accompany a
logarithm $\ln m_q$, we omit it for simplicity as the $m_q^3$ term itself
is a minor correction.
The subtracted condensate $\langle\bar{q}q\rangle^{\rm (subt)}$ is
then free from power divergences, but could still contain
non-divergent $m_q$ dependence, such as the chiral logarithm.

\begin{figure}
 \includegraphics[width=9.0cm,clip]{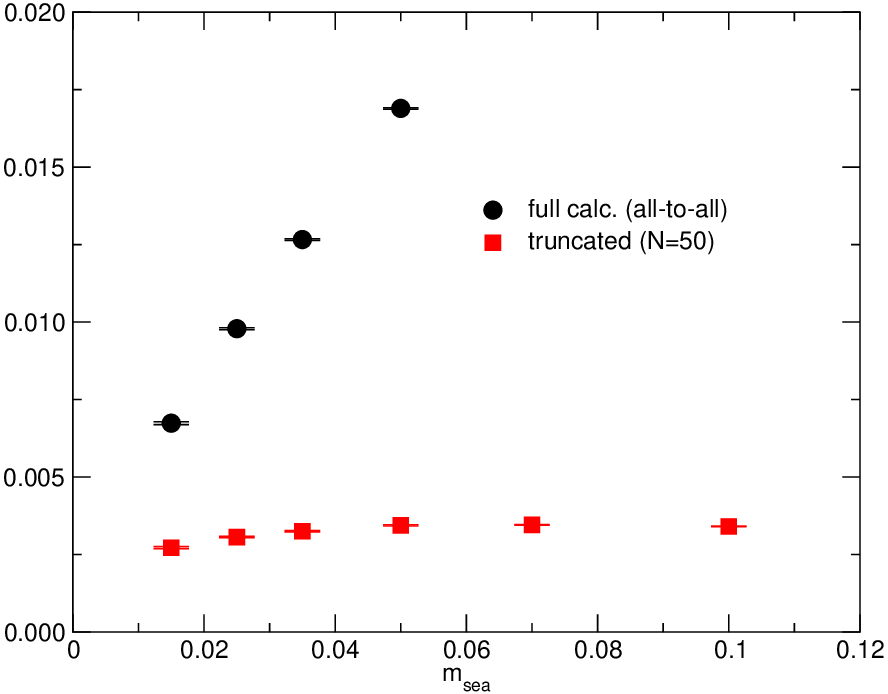}
 \caption{
   Comparison between $\vev{\bar{q}q}$ obtained from all-to-all
   propagator (circles) and $\vev{\bar{q}q}^{(N=50)}$ (squares) as a
   function of $m_{\rm sea}=m_q$.
   The data obtained on NF2p are shown.
 }
 \label{PBPcmpr}
\end{figure}

In Figure~\ref{PBPcmpr}, we compare a ``full'' calculation of 
$\vev{\bar{q}{q}}$ (circles) corresponding to $N=12L_s^3L_t$
and $\vev{\bar{q}q}^{(N=50)}$.
The ``full'' calculation contains the contributions of all eigenmodes
which are evaluated by a stochastic method. 
(Our set-up is explained in \cite{JLQCD:2009qn}.)
The data on the NF2p lattice with sea and valence quark masses set
equal are shown as an example.
The results clearly show that the divergent term $m_q/a^2$ in
(\ref{pbpN}) dominates the ``full'' condensate, and it seems difficult
to extract $\langle\bar{q}q\rangle^{\rm (subt)}$ from this data alone.
The truncated condensate, on the other hand, does not have that strong
$m_q$ dependence, but still both the $m_q/a^2$ and $m_q^3$ terms are
visible. 

\begin{figure}[tbp]
 \includegraphics[width=8.0cm,clip]{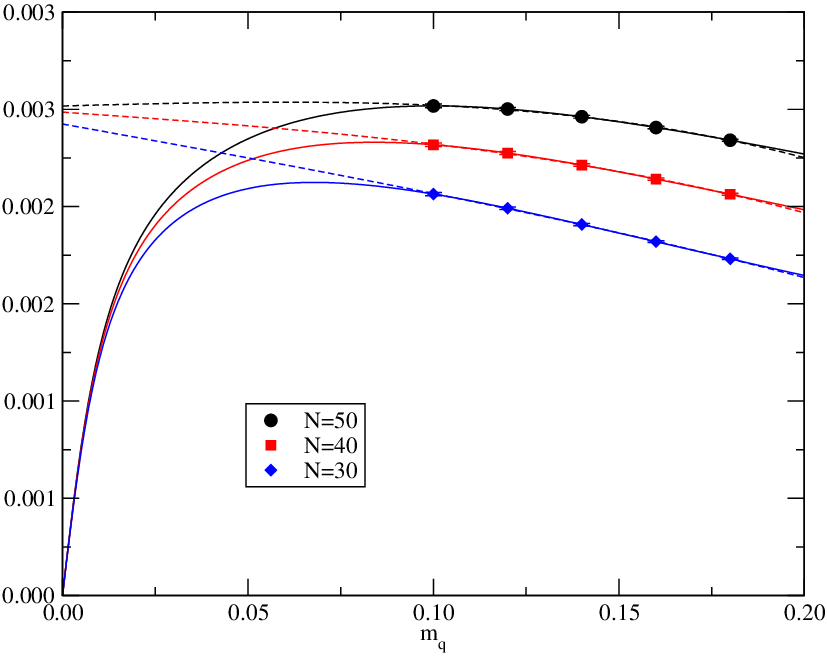}
 \includegraphics[width=8.0cm,clip]{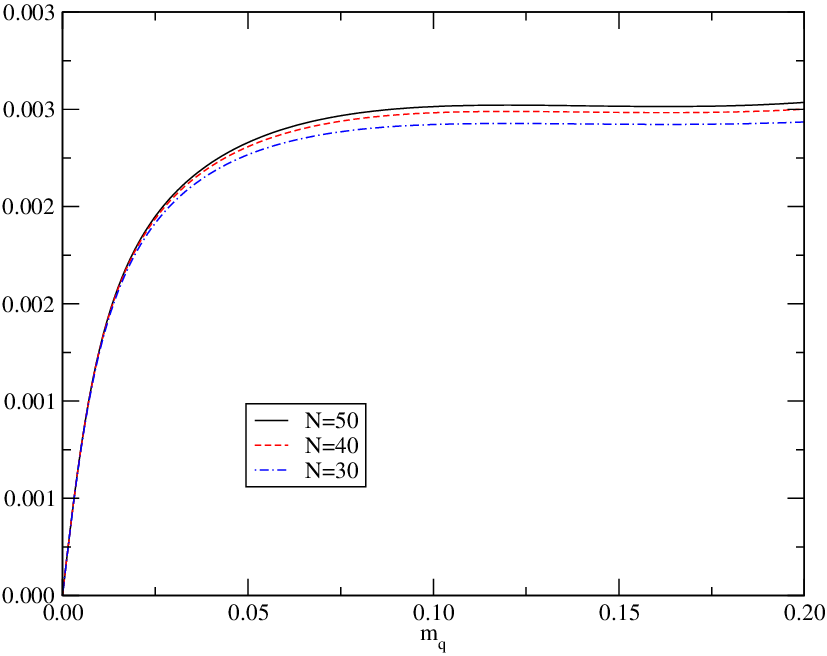}
 \caption{
   Subtraction of the power divergence in the chiral condensate.
   The left panel shows $\vev{\bar{q}q}^{(N)}$ calculated on the
   NF2$\epsilon$ lattice as a function of the valence quark mass $m_q$
   (solid curves). 
   The number of low-modes included are $N$ = 50, 40 and 30 from top to
   bottom.
   The solid curves are data constructed from calculated eigenvalues,
   and the points with error bars are representative points used in
   our fit.
   The fit curves according to  (\ref{pbpN}) are shown by dashed curves. 
   The right panel represents the subtracted condensate
   $\vev{\bar{q}q}^{\rm (subt)}$. 
 }
 \label{PBPsubt_NF2e}
\end{figure}

\begin{figure}[tbp]
 \includegraphics[width=8.0cm,clip]{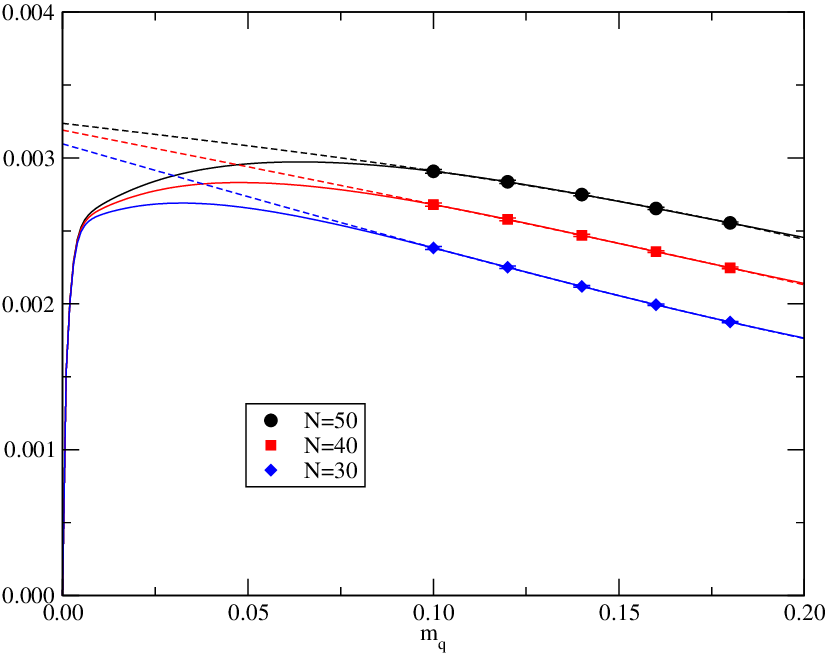}
 \includegraphics[width=8.0cm,clip]{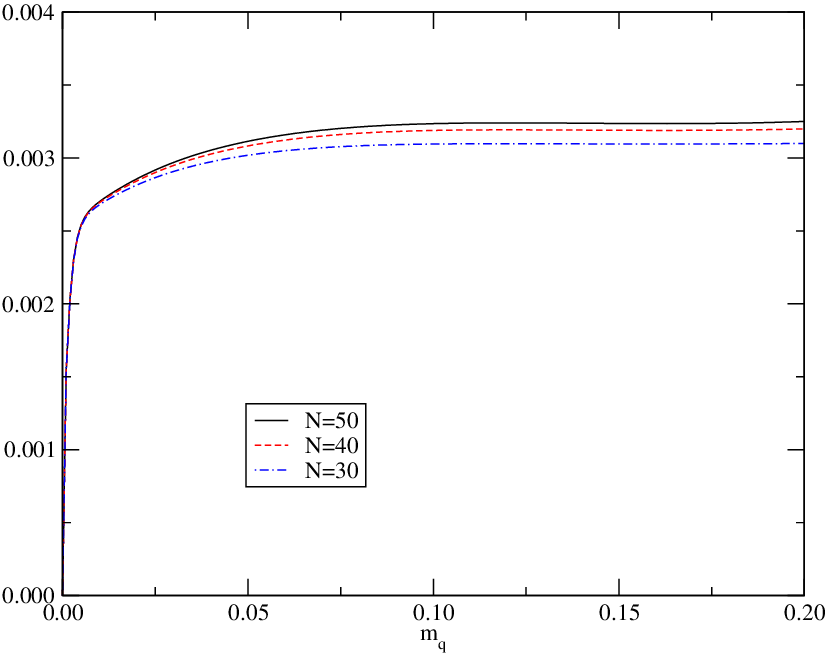}
 \caption{
   Same as Figure~\ref{PBPsubt_NF2e} but for NF2p at $m_{\rm sea}= 0.015$.
 }
 \label{PBPsubt_NF2p}
\end{figure}

\begin{figure}[tbp]
 \includegraphics[width=8.0cm,clip]{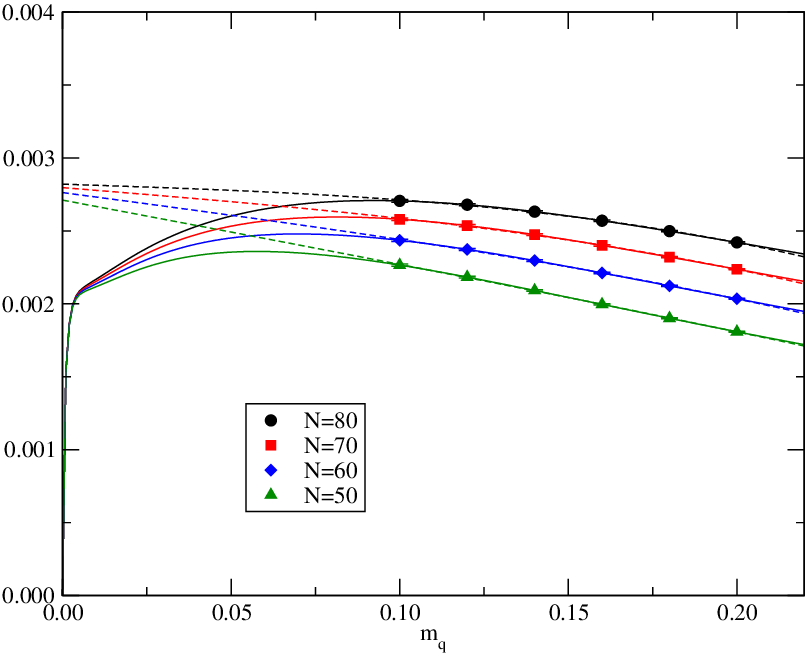}
 \includegraphics[width=8.0cm,clip]{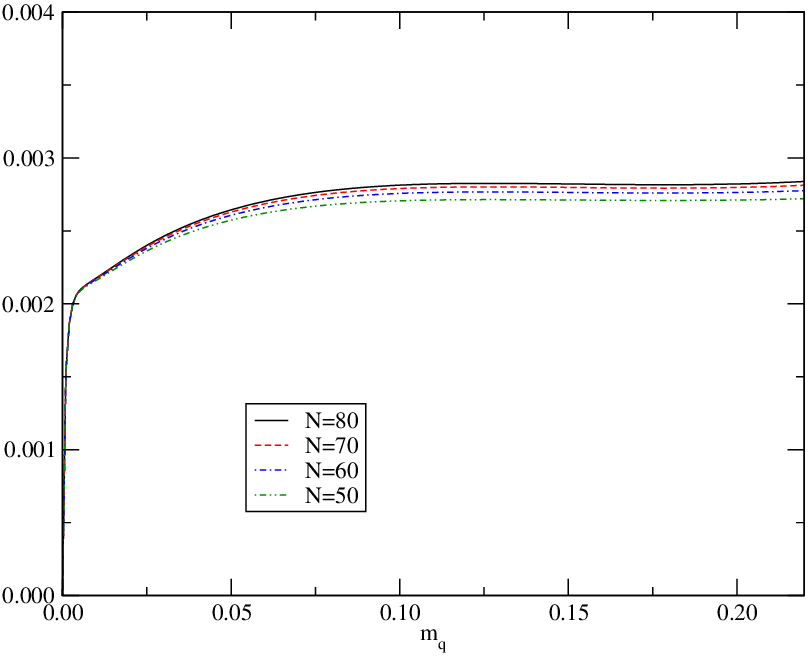}\\
 \includegraphics[width=8.0cm,clip]{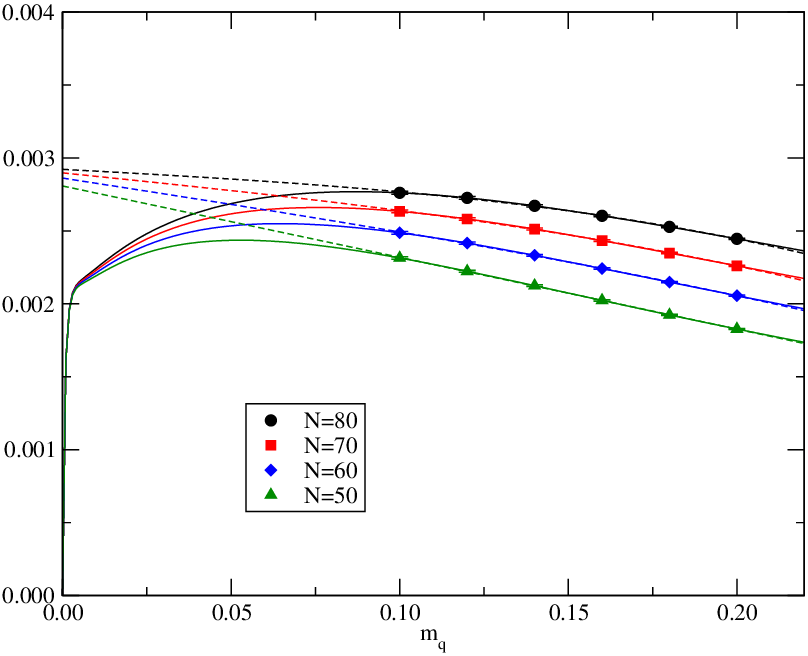}
 \includegraphics[width=8.0cm,clip]{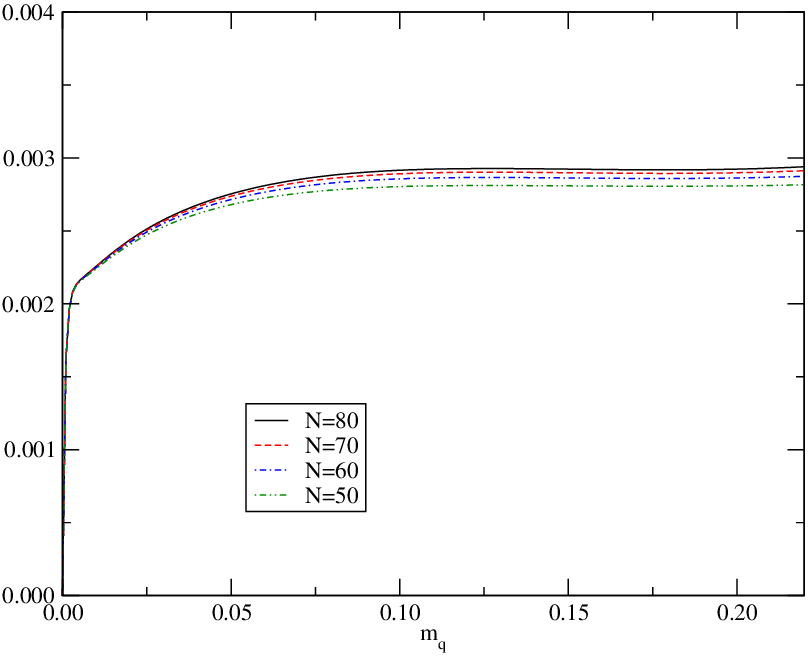}
 \caption{
   Same as Figure~\ref{PBPsubt_NF2e} but for NF3p-a (top) and NF3p-b
   (bottom) at $m_{ud}= 0.015$.
   The number of low-modes included are $N$ = 80, 70, 60 and 50 from
   top to bottom.
 }
 \label{PBPsubt_NF3}
\end{figure}

We now try to extract the non-divergent term
$\langle\bar{q}q\rangle^{\rm (subt)}$ using (\ref{pbpN}).
In Figures~\ref{PBPsubt_NF2e}--\ref{PBPsubt_NF3} (left panel) we plot
the truncated condensate $\vev{\bar{q}q}^{(N)}$ as a function of the 
valence quark mass with three or four different values of $N$.
The data are shown for individual lattice ensembles (NF2$\epsilon$, NF2p, NF3p-a
and NF3p-b); except for NF2$\epsilon$ the results at the lightest sea quark are
shown as an example.
The truncated condensate can be constructed at arbitrary values of the
valence quark mass $m_q$ without extra computational costs.
In order to see the ultraviolet behavior, we plot in the mass region
up to $m_q$ = 0.20, which is twice larger than the largest simulated
sea quark mass.

When we fit the lattice data to (\ref{pbpN}), we take five or six
representative points of $m_q$ in the range 
$0.10 \le m_q \le 0.18$ for NF2$\epsilon$ and NF2p or
$0.10 \le m_q \le 0.20$ for NF3p-a and NF3p-b.
The upper limit is chosen such that $|\hat{\lambda}_N|>m_q$ for the
given $N$.
Otherwise we do not expect the ultraviolet behavior (\ref{pbpN}).
In this rather heavy mass region, we do not expect additional mass
dependence from the infrared origin, and we simply set
$-\vev{\bar{q}q}^{\rm (subt)}=\Sigma$ with $\Sigma$ a constant.

The fit results are shown in
Figures~\ref{PBPsubt_NF2e}--\ref{PBPsubt_NF3} (left panel) by dashed
curves. 
In each right panel of these figures,
the subtracted condensates
$\vev{\bar{q}q}^{\rm (subt)}= \vev{\bar{q}q}^{(N)}
- c_1^{(N)}/a^2 m_q -c_2^{(N)}m_q^3$ 
for the same choices of $N$ as the left panel are shown.
We observe that the subtracted condensate depends on $N$ very mildly to
 a few \% order.
It implies that our subsequent analysis using $\vev{\bar{q}q}^{(N)}$ 
with the maximal $N$ may contain a small systematic error due to the truncation
of $N$. We discuss this point and estimate the error in Section~\ref{Results}.

\begin{figure}
 \includegraphics[width=8cm,clip]{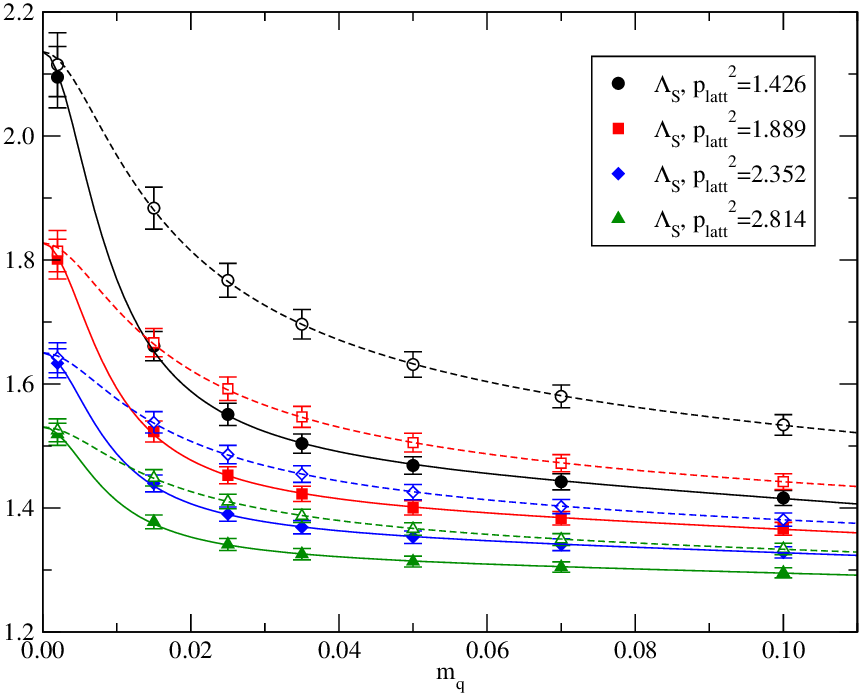}
 \includegraphics[width=8cm,clip]{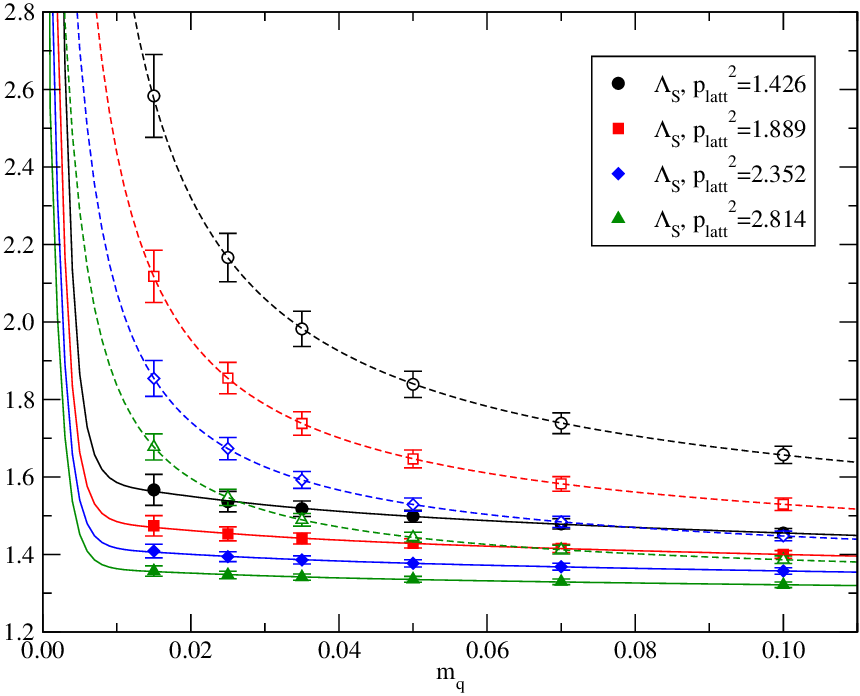}
 \includegraphics[width=8cm,clip]{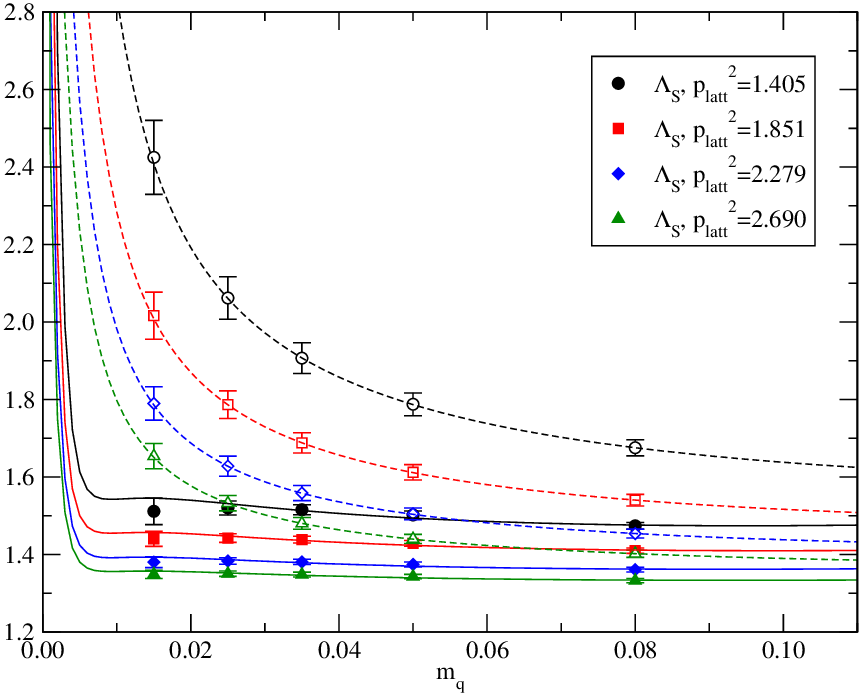}
 \includegraphics[width=8cm,clip]{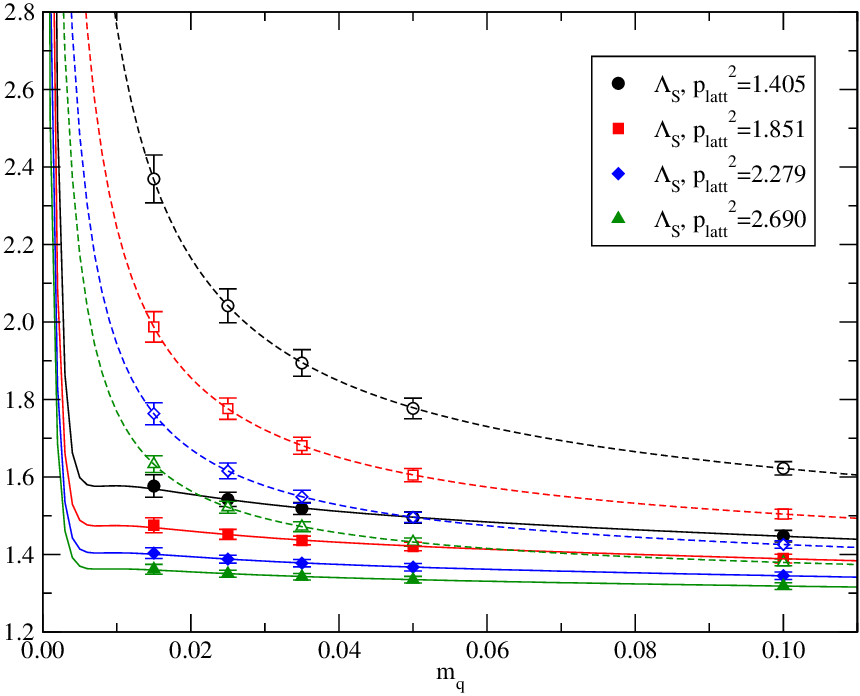}
 \caption{
   Vertex functions $\Lambda^{\rm latt}_S(p_{\rm latt})$ 
   (filled symbols) and $\Lambda^{\rm latt}_P(p_{\rm latt})$ 
   (open symbols).
   The data are shown for the NF2$\epsilon$ (upper left), NF2p (upper
   right), NF3p-a (lower left) and NF3p-b (lower right) lattices as
   functions of $m_q$. 
   In each panel, data at selected values of $p_{\rm latt}^2$ and
   their fit curves are presented. 
   For NF2p, NF3p-a and NF3p-b, data with a fixed sea quark mass 
   $m_{\rm sea} (m_{ud}) = 0.015$ are plotted as an example.
 }
 \label{SPvm}
\end{figure}

We use $\vev{\bar{q}q}^{\rm (subt)}$ thus obtained at each sea quark mass
as a function of the valence quark mass in the analysis of the scalar and
pseudo-scalar vertex functions, (\ref{LmdS}) and (\ref{LmdP}), respectively. 
The valence quark mass dependence of 
$\Lambda^{\rm latt}_S(p_{\rm latt})$ and  
$\Lambda^{\rm latt}_P(p_{\rm latt})$ at four representative values of
$p_{\rm latt}^2$ are plotted in Figure~\ref{SPvm}.
We find that both the scalar (filled symbols) and pseudo-scalar (open
symbols) vertices are nicely described by the fit curves according to 
(\ref{LmdS}) and (\ref{LmdP}) supplemented by the measured 
$\vev{\bar{q}q}^{\rm (subt)}$.
In particular, as seen near the chiral limit of the NF2$\epsilon$ data,
the fit curves precisely reproduce 
the agreement of $\Lambda^{\rm latt}_S(p_{\rm latt})$ and  
$\Lambda^{\rm latt}_P(p_{\rm latt})$ in the $\epsilon$-regime, which is 
not expected when $\vev{\bar{q}q}$ is treated as a mass-independent constant.

In addition to (\ref{LmdS}) and (\ref{LmdP}), quadratic mass-dependence 
is possible for the vertex functions $\Lambda^{\rm latt}_S(p_{\rm latt})$ and  
$\Lambda^{\rm latt}_P(p_{\rm latt})$:
\begin{eqnarray}
 \Lambda_S(p_{\rm latt}) 
  &=& \frac{C}{p^2} \frac{\partial \vev{\bar{q}q}}{\partial m_q}
  +Z_qZ_m + B_S m_q^2,\\
 \Lambda_P(p_{\rm latt}) 
  &=& \frac{C}{p^2} \frac{\vev{\bar{q}q}}{m_q} +Z_qZ_m + B_P m_q^2.
\end{eqnarray}
From a combined fit of the valence quark mass dependence,
we obtain the parameters $C$, $Z_qZ_m$, $B_P$ and $B_S$ at each value
of $p_{\rm latt}^2$.
Numerical results are listed in
Tables~\ref{SPFIT_NF2e}--\ref{SPFIT_NF3p-b}
for each sea quark masses of the NF2$\epsilon$, NF2p, NF3p-a and NF3p-b
lattices.
We find that the values of $C$ depend on $p^2$ only mildly, which
is consistent with the logarithmic dependence through $\alpha_s$.

\begin{table}
 \begin{tabular}{cccccc}
  \hline\hline
   $p_{\rm latt}^2$ & $C$ & $Z_qZ_m$ & $B_S$ & $B_P$ & $\chi^2/$dof\\
  \hline
  1.426& 5.92(40) & 1.443(16) & $-$3.10(90)& $-$1.35(34)& 0.025\\
  1.889& 5.01(32) & 1.384(13) & $-$2.08(62)& $-$0.85(25)& 0.017\\
  2.352& 4.32(28) & 1.342(11) & $-$1.60(48)& $-$0.76(22)& 0.016\\
  2.814& 3.78(26) & 1.3060(92)& $-$1.23(36)& $-$0.65(15)& 0.017\\
  \hline\hline
  \end{tabular}
 \caption{
   Parameters in the simultaneous fit of $\Lambda_S^{\rm latt}$ and
   $\Lambda_P^{\rm latt}$ for NF2$\epsilon$.
   Results at representative values of lattice momenta are listed.
 }
 \label{SPFIT_NF2e}
\end{table}

\begin{table}
 \begin{tabular}{ccccccc}
  \hline\hline
  $m_{\rm sea}$& $p_{\rm latt}^2$& $C$& $Z_qZ_m$& $B_S$& $B_P$& $\chi^2/$dof\\
  \hline
  $0.015$
  &1.426&\ \ 8.48(83) &\ \ 1.472(22) &\ \ $-$2.0(1.9)&\ \ $-$0.80(91)&\ \ 0.004\\
  &1.889&\ \ 7.16(68) &\ \ 1.412(16) &\ \ $-$1.5(1.3)&\ \ $-$0.57(66)&\ \ 0.005\\
  &2.352&\ \ 6.19(59) &\ \ 1.366(11) &\ \ $-$1.01(95)&\ \ $-$0.35(46)&\ \ 0.005\\
  &2.814&\ \ 5.31(51) &\ \ 1.3271(89)&\ \ $-$0.65(75)&\ \ $-$0.18(38)&\ \ 0.004\\
  \hline
  $0.025$
  &1.426&\ \ 7.43(39) &\ \ 1.497(20) &\ \ $-$5.3(1.1)&\ \ $-$2.21(49)&\ \ 0.139\\
  &1.889&\ \ 6.29(32) &\ \ 1.429(15) &\ \ $-$3.53(76)&\ \ $-$1.33(35)&\ \ 0.084\\
  &2.352&\ \ 5.43(26) &\ \ 1.380(13) &\ \ $-$2.66(56)&\ \ $-$1.07(25)&\ \ 0.071\\
  &2.814&\ \ 4.61(23) &\ \ 1.341(10) &\ \ $-$2.18(43)&\ \ $-$0.96(23)&\ \ 0.094\\
  \hline
  $0.035$
  &1.426&\ \ 8.57(91) &\ \ 1.446(24) &\ \ $-$0.2(2.0)&\ \ $-$0.1(1.1)&\ \ 0.190\\
  &1.889&\ \ 7.37(77) &\ \ 1.392(17) &\ \ \ 0.1(1.4)&\ \ \ 0.11(71)  &\ \ 0.179\\
  &2.352&\ \ 6.23(61) &\ \ 1.353(13) &\ \ $-$0.13(92)&\ \ $-$0.02(48)&\ \ 0.133\\
  &2.814&\ \ 5.40(56) &\ \ 1.317(11) &\ \ \ 0.00(76)&\ \ \ 0.09(43)  &\ \ 0.129\\
  \hline
  $0.050$
  &1.426&\ \ 9.00(58) &\ \ 1.469(21) &\ \ $-$1.3(1.5)&\ \ $-$0.86(63)&\ \ 0.021\\
  &1.889&\ \ 7.61(49) &\ \ 1.419(15) &\ \ $-$1.1(1.1)&\ \ $-$0.54(48)&\ \ 0.014\\
  &2.352&\ \ 6.54(42) &\ \ 1.377(12) &\ \ $-$0.95(81)&\ \ $-$0.59(36)&\ \ 0.019\\
  &2.814&\ \ 5.60(37) &\ \ 1.340(10) &\ \ $-$0.89(63)&\ \ $-$0.61(31)&\ \ 0.047\\
  \hline
  $0.070$
  &1.426&\ \ 7.33(54) &\ \ 1.476(19) &\ \ $-$5.1(1.4)&\ \ $-$2.19(63)&\ \ 0.303\\
  &1.889&\ \ 6.23(45) &\ \ 1.412(14) &\ \ $-$3.56(99)&\ \ $-$1.50(45)&\ \ 0.267\\
  &2.352&\ \ 5.36(40) &\ \ 1.363(11) &\ \ $-$2.61(78)&\ \ $-$1.03(38)&\ \ 0.228\\
  &2.814&\ \ 4.56(34) &\ \ 1.3262(93)&\ \ $-$2.23(58)&\ \ $-$0.94(30)&\ \ 0.249\\
  \hline
  $0.100$
  &1.426&\ \ 8.24(61) &\ \ 1.444(19) &\ \ $-$1.9(1.5)&\ \ $-$1.09(62)&\ \ 0.008\\
  &1.889&\ \ 6.99(53) &\ \ 1.394(14) &\ \ $-$1.4(1.1)&\ \ $-$0.81(46)&\ \ 0.014\\
  &2.352&\ \ 6.04(47) &\ \ 1.351(12) &\ \ $-$0.92(88)&\ \ $-$0.51(42)&\ \ 0.003\\
  &2.814&\ \ 5.18(42) &\ \ 1.316(10) &\ \ $-$0.72(74)&\ \ $-$0.32(39)&\ \ 0.004\\
  \hline\hline
 \end{tabular}
 \caption{Same as Table~\ref{SPFIT_NF2e} but for NF2p.}
 \label{SPFIT_NF2p}
\end{table}

\begin{table}
 \begin{tabular}{ccccccc}
  \hline\hline
  $m_{ud}$ &\ \ $p_{\rm latt}^2$ &\ \ $C$ &\ \ $Z_qZ_m$ 
   &\ \ $B_S$ &\ \ $B_P$ &\ \ $\chi^2/$dof\\
  \hline
  $0.015$
  &1.405&\ \ 9.00(98)&\ \ 1.445(21) &\ \ \ 2.3(2.8)&\ \ \ 1.3(1.5)&\ \ 0.345\\
  &1.851&\ \ 7.58(82)&\ \ 1.393(14) &\ \ \ 1.3(1.9)&\ \ \ 0.9(1.1)&\ \ 0.248\\
  &2.279&\ \ 6.65(72)&\ \ 1.347(10) &\ \ \ 1.2(1.5)&\ \ \ 0.87(86)&\ \ 0.276\\
  &2.690&\ \ 5.86(62)&\ \ 1.3228(83)&\ \ \ 0.8(1.1)&\ \ \ 0.55(58)&\ \ 0.194\\
  \hline
  $0.025$
  &1.405&\ \ 7.18(52)&\ \ 1.477(14) &\ \ $-$3.5(1.4)&\ \ $-$1.59(60)&\ \ 0.046\\
  &1.851&\ \ 6.14(44)&\ \ 1.418(11) &\ \ $-$2.77(94)&\ \ $-$1.41(42)&\ \ 0.056\\
  &2.279&\ \ 5.25(37)&\ \ 1.3661(93)&\ \ $-$2.07(72)&\ \ $-$1.03(35)&\ \ 0.046\\
  &2.690&\ \ 4.64(32)&\ \ 1.3379(78)&\ \ $-$1.86(53)&\ \ $-$0.92(26)&\ \ 0.065\\
  \hline
  $0.035$
  &1.405&\ \ 7.21(54)&\ \ 1.495(19) &\ \ $-$3.3(2.0)&\ \ $-$1.90(81)&\ \ 0.018\\
  &1.851&\ \ 6.19(46)&\ \ 1.427(15) &\ \ $-$2.1(1.5)&\ \ $-$1.17(58)&\ \ 0.018\\
  &2.279&\ \ 5.36(40)&\ \ 1.376(12) &\ \ $-$1.6(1.1)&\ \ $-$0.91(48)&\ \ 0.015\\
  &2.690&\ \ 4.75(36)&\ \ 1.3448(97)&\ \ $-$1.20(88)&\ \ $-$0.58(38)&\ \ 0.010\\
  \hline
  $0.050$
  &1.405&\ \ 7.88(56)&\ \ 1.444(16) &\ \ $-$0.2(1.9)&\ \ $-$0.34(82)&\ \ 0.098\\
  &1.851&\ \ 6.80(52)&\ \ 1.391(12) &\ \ \  0.0(1.5)&\ \ $-$0.03(72)&\ \ 0.099\\
  &2.279&\ \ 5.82(44)&\ \ 1.3464(92)&\ \ \  0.0(1.1)&\ \ \  0.14(52)&\ \ 0.086\\
  &2.690&\ \ 5.27(43)&\ \ 1.3216(80)&\ \ \  0.08(98)&\ \ \  0.10(48)&\ \ 0.075\\
  \hline
  $0.080$
  &1.405&\ \ 9.24(67)&\ \ 1.449(19) &\ \ $-$0.7(2.6)&\ \ $-$9.35(78)&\ \ 0.211\\
  &1.851&\ \ 8.00(59)&\ \ 1.395(13) &\ \ $-$0.2(1.9)&\ \ $-$5.81(59)&\ \ 0.226\\
  &2.279&\ \ 6.81(50)&\ \ 1.352(10) &\ \ $-$0.1(1.4)&\ \ $-$3.93(47)&\ \ 0.189\\
  &2.690&\ \ 6.06(44)&\ \ 1.3296(86)&\ \ $-$0.2(1.1)&\ \ $-$3.01(38)&\ \ 0.138\\
  \hline\hline
 \end{tabular}
 \caption{Same as Table~\ref{SPFIT_NF2e} but for NF3p-a.}
 \label{SPFIT_NF3p-a}
\end{table}

\begin{table}
 \begin{tabular}{ccccccc}
  \hline\hline
  $m_{ud}$ &\ \ $p_{\rm latt}^2$ &\ \ $C$ &\ \ $Z_qZ_m$
   &\ \ $B_S$ &\ \ $B_P$ &\ \ $\chi^2/$dof\\
  \hline
  $0.015$
  &1.405&\ \ 8.08(59)&\ \ 1.464(13) &\ \ $-$2.30(94)&\ \ $-$0.99(55)&\ \ 0.020\\
  &1.851&\ \ 6.92(49)&\ \ 1.4009(98)&\ \ $-$1.57(63)&\ \ $-$0.57(35)&\ \ 0.029\\
  &2.279&\ \ 5.97(42)&\ \ 1.3527(78)&\ \ $-$1.04(50)&\ \ $-$0.34(29)&\ \ 0.010\\
  &2.690&\ \ 5.30(37)&\ \ 1.3239(68)&\ \ $-$0.76(41)&\ \ $-$0.19(22)&\ \ 0.007\\
  \hline
  $0.025$
  &1.405&\ \ 7.04(42)&\ \ 1.4699(93)&\ \ $-$2.61(78)&\ \ $-$1.21(42)&\ \ 0.012\\
  &1.851&\ \ 6.01(36)&\ \ 1.4101(75)&\ \ $-$1.88(56)&\ \ $-$0.83(31)&\ \ 0.004\\
  &2.279&\ \ 5.19(32)&\ \ 1.3595(60)&\ \ $-$1.34(43)&\ \ $-$0.54(22)&\ \ 0.003\\
  &2.690&\ \ 4.65(29)&\ \ 1.3321(52)&\ \ $-$1.11(34)&\ \ $-$0.51(18)&\ \ 0.005\\
  \hline
  $0.035$
  &1.405&\ \ 7.14(50)&\ \ 1.475(11) &\ \ $-$3.44(78)&\ \ $-$1.39(43)&\ \ 0.146\\
  &1.851&\ \ 6.09(41)&\ \ 1.4126(85)&\ \ $-$2.47(54)&\ \ $-$1.07(27)&\ \ 0.087\\
  &2.279&\ \ 5.26(35)&\ \ 1.3624(68)&\ \ $-$1.84(41)&\ \ $-$0.77(22)&\ \ 0.095\\
  &2.690&\ \ 4.67(31)&\ \ 1.3320(58)&\ \ $-$1.31(32)&\ \ $-$0.48(18)&\ \ 0.059\\
  \hline
  $0.050$
  &1.405&\ \ 8.93(56)&\ \ 1.423(15) &\ \ $-$0.1(1.3)&\ \ \  0.03(55)&\ \ 0.295\\
  &1.851&\ \ 7.59(47)&\ \ 1.3755(99)&\ \ $-$0.09(84)&\ \ \  0.03(37)&\ \ 0.325\\
  &2.279&\ \ 6.52(41)&\ \ 1.3363(77)&\ \ $-$0.16(63)&\ \ $-$0.01(29)&\ \ 0.200\\
  &2.690&\ \ 5.85(36)&\ \ 1.3115(65)&\ \ $-$0.04(50)&\ \ \  0.02(22)&\ \ 0.240\\
  \hline
  $0.100$
  &1.405&\ \ 8.68(56)&\ \ 1.468(16) &\ \ $-$1.7(1.2)&\ \ $-$0.61(60)&\ \ 0.123\\
  &1.851&\ \ 7.34(48)&\ \ 1.412(12) &\ \ $-$1.21(91)&\ \ $-$0.49(46)&\ \ 0.113\\
  &2.279&\ \ 6.26(40)&\ \ 1.3650(87)&\ \ $-$1.00(66)&\ \ $-$0.41(29)&\ \ 0.072\\
  &2.690&\ \ 5.61(35)&\ \ 1.3353(75)&\ \ $-$0.59(54)&\ \ $-$0.24(26)&\ \ 0.104\\
  \hline\hline
 \end{tabular}
 \caption{Same as Table~\ref{SPFIT_NF2e} but for NF3p-b.}
 \label{SPFIT_NF3p-b}
\end{table}

\subsection{Renormalization constants}~\label{Results}

From the fits described in the previous subsections, we obtain the
numerical results for $Z_q$ and $Z_qZ_m=Z_q/Z_S$ for each available
values of $p_{\rm latt}$.
From a similar analysis, we also obtain $\Lambda_T=Z_qZ_T^{-1}$, which
does not depend on the quark mass significantly.
We combine these results with $Z_q(\mu)$ to obtain
$Z_m^{\rm RI/MOM}(\mu)=1/Z_S^{\rm RI/MOM}(\mu)$ and 
$Z_T^{\rm RI/MOM}(\mu)$ as functions of the renormalization scale $\mu$. 

\begin{figure}[tbp]
 \includegraphics[width=8cm,clip]{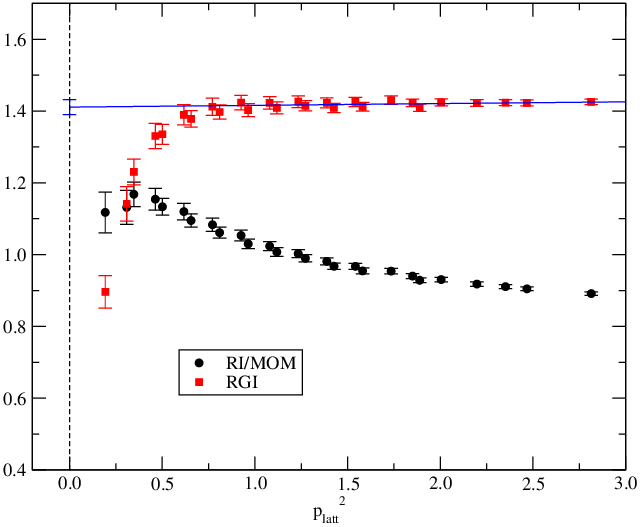}
 \includegraphics[width=8cm,clip]{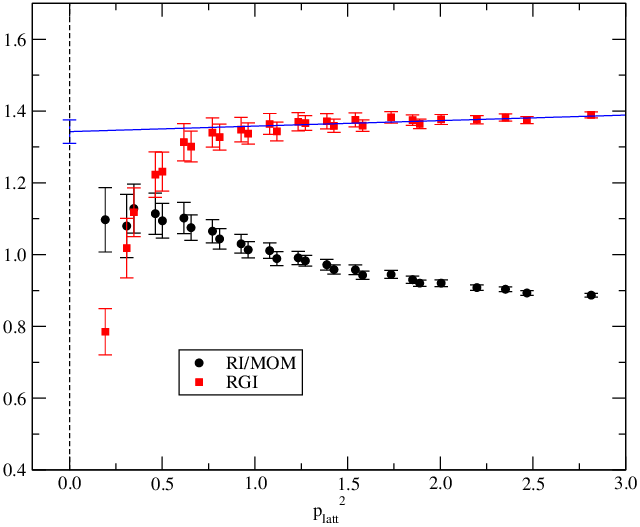}
 \includegraphics[width=8cm,clip]{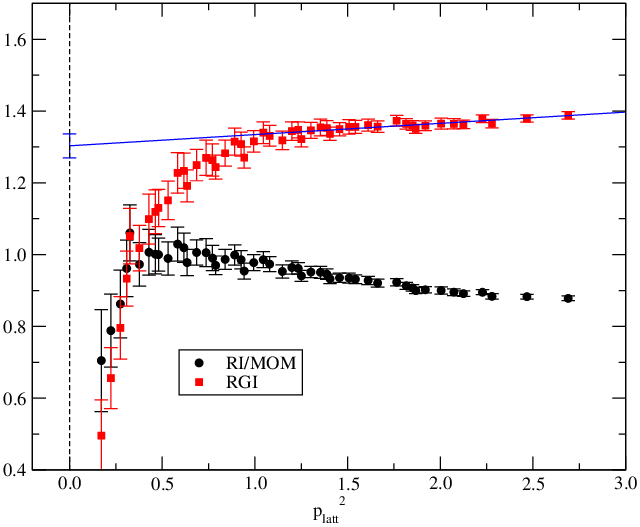}
 \includegraphics[width=8cm,clip]{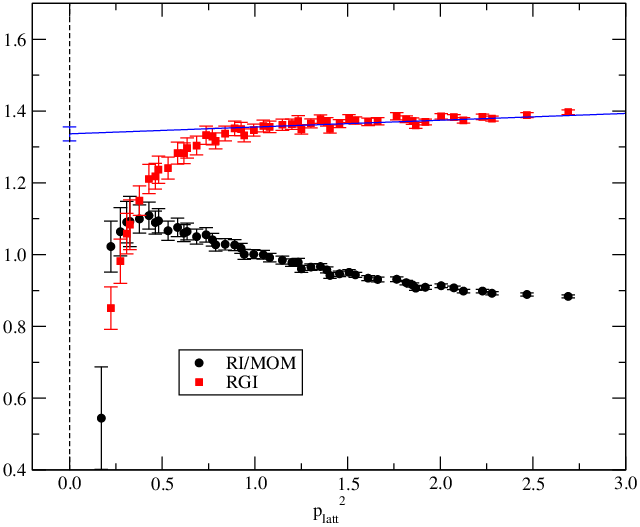}
 \caption{
   Renormalization factors for the quark mass $Z_m$ 
   in RI/MOM scheme (circles) and their RGI values (squares) as
   functions of $p_{\rm latt}^2$ for NF2$\epsilon$ (upper left), NF2p
   (upper right), NF3p-a (lower left), and NF3p-b (lower right).
   Results of the linear extrapolation to the $(p_{\rm latt})^2\to 0$
   limit of $Z_m^{\rm RGI}$ are shown as well. 
   For NF2p, NF3p-a and NF3p-b, data with a fixed sea quark mass 
   $m_{\rm sea} (m_{ud}) = 0.015$ are plotted as an example.
 }
 \label{Matching_mq}
\end{figure}

\begin{figure}[tbp]
 \includegraphics[width=8cm,clip]{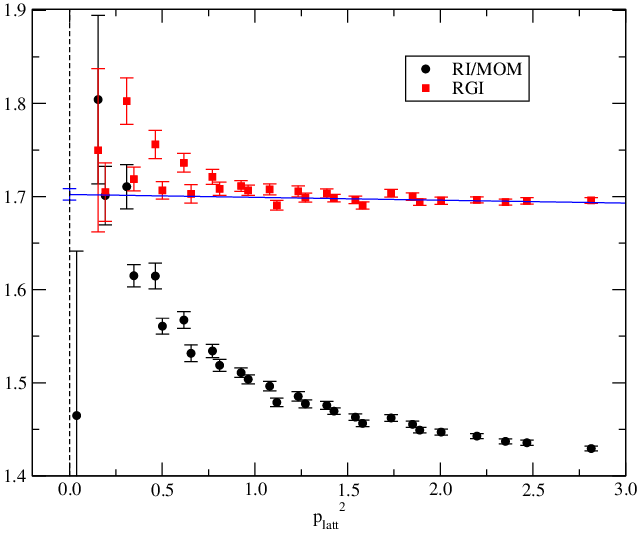}
 \includegraphics[width=8cm,clip]{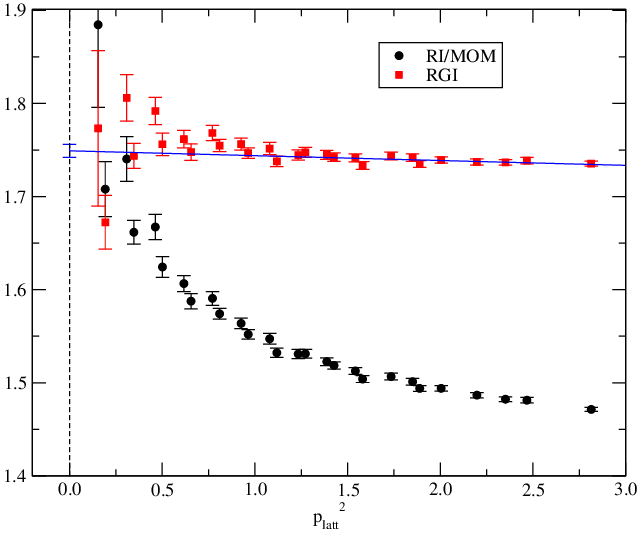}
 \includegraphics[width=8cm,clip]{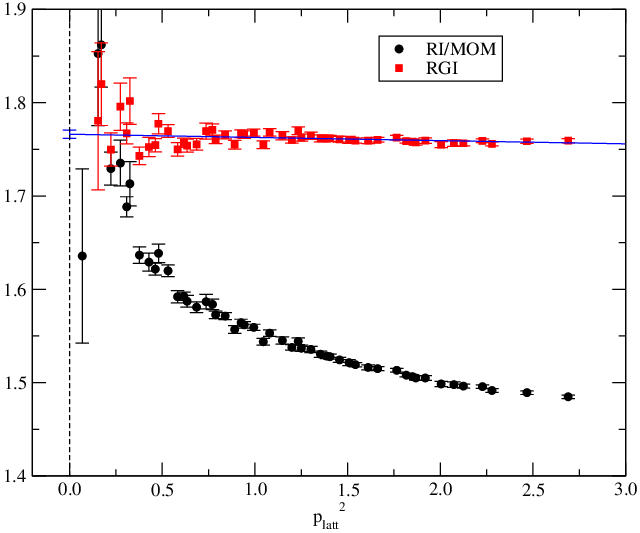}
 \includegraphics[width=8cm,clip]{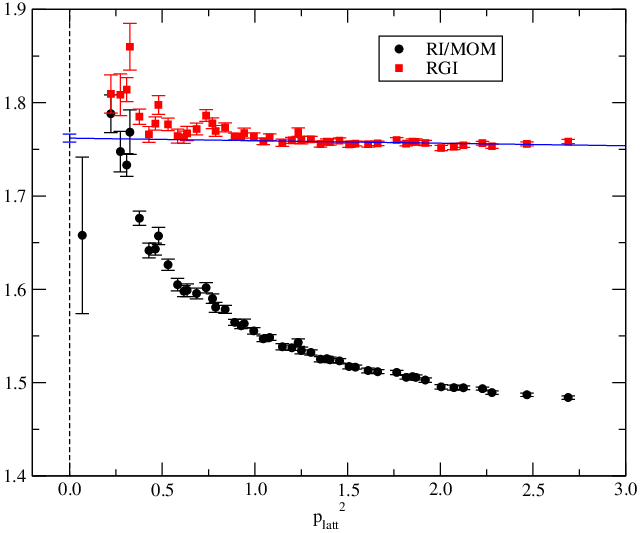}
 \caption{
   Same as Figure~\ref{Matching_mq} but for the tensor current
   renormalization factor $Z_T$.
 }
 \label{Matching_T}
\end{figure}

The results are plotted in Figures~\ref{Matching_mq} and
\ref{Matching_T} for $Z_m$ and $Z_T$, respectively.
Filled black symbols representing the numerical data for the RI/MOM
scheme clearly show a scale (or $p_{\rm latt}^2$) dependence.
This dependence can partly be absorbed by perturbatively calculated
matching factor $w_{\cal O}^{\rm RI/MOM}(\mu)$ ($\cal O$ = $m$ or $T$)
to the RGI values as in the case of $Z_q$. 
The perturbative results for $w_{\cal O}^{\rm RI/MOM}(\mu)$ and 
$w_{\cal O}^{\ovl{\rm MS}}(\mu)$ are summarized in
Appendix~\ref{matching}. 

The numerical data for 
$Z_{\cal O}^{\rm RGI} = 
Z_{\cal O}^{\rm RI/MOM}(p_{\rm latt})/w_{\cal O}^{\rm RI/MOM}(p_{\rm latt})$
are also shown in Figures~\ref{Matching_mq} and \ref{Matching_T}.
We find that the scale dependence is largely absorbed at least above
$(p_{\rm latt})^2\simeq 1$, as expected.
Below $(p_{\rm latt})^2\simeq 1$ the perturbative estimate of
$w_{\cal O}^{\rm RI/MOM}(p)$ becomes less precise even though three-
or four-loop calculations are used.
Remaining scale dependence above $(p_{\rm latt})^2\simeq 1$ is
ascribed to the discretization effect of $O(a^2)$.
We therefore extrapolate the data for $Z_{\cal O}^{\rm RGI}$ above
$(p_{\rm latt})^2=1$ to the vanishing $(p_{\rm latt})^2$ limit assuming a
linear dependence on $(p_{\rm latt})^2$, which is shown by solid lines
in Figures~\ref{Matching_mq} and \ref{Matching_T}.

\begin{table}[tbp]
  \begin{tabular}{rcccc}
   \hline\hline
   $m_{\rm sea}(m_{ud})$
   &\ \ \ \ $Z_S^{\rm RGI}$\ \ \ \ &\ \ \ \ $Z_m^{\rm RGI}$\ \ \ \ 
   & $Z_S^{\ovl{\rm MS}}(2\ {\rm GeV})$ & $Z_m^{\ovl{\rm MS}}(2\ {\rm GeV})$\\
   \hline
   NF2$\epsilon$,\  
   0.002\ \ &0.709(11) &  1.411(21)&   1.205(18)&   0.830(12)\\
   \hline\hline
   NF2p,\ 
   0.015\ \ &0.743(18) &  1.345(33)&   1.263(30)&   0.791(19)\\
   0.025\ \ &0.719(12) &  1.390(24)&   1.223(21)&   0.818(14)\\
   0.035\ \ &0.764(20) &  1.308(37)&   1.298(35)&   0.769(22)\\
   0.050\ \ &0.746(16) &  1.339(29)&   1.268(27)&   0.788(17)\\
   0.070\ \ &0.726(14) &  1.378(26)&   1.234(23)&   0.810(15)\\
   0.100\ \ &0.761(14) &  1.313(26)&   1.293(24)&   0.772(15)\\
   \hline
   0.000\ \ & 0.7309(87)& 1.366(16) & 1.243(15) &   0.8035(97)\\
   ($\chi^2/$dof&  1.15 &  1.15     &   1.15    &   1.15)   \\
   \hline\hline
   NF3p-a,\ 
   0.015\ \ &0.766(19) & 1.303(34)&  1.296(32)& 0.770(20) \\ 
   0.025\ \ &0.734(10) & 1.362(20)&  1.242(17)& 0.805(12) \\
   0.035\ \ &0.721(17) & 1.386(33)&  1.221(29)& 0.819(19) \\
   0.050\ \ &0.757(15) & 1.320(27)&  1.281(25)& 0.780(16) \\
   0.080\ \ &0.765(17) & 1.304(31)&  1.296(29)& 0.770(18) \\
   \hline
   NF3p-b,\ 
   0.015\ \ &0.748(11) & 1.337(19)&  1.265(18)& 0.790(11) \\
   0.025\ \ &0.7354(62)& 1.359(12)&  1.245(11)& 0.8031(70)\\
   0.035\ \ &0.7311(87)& 1.368(17)&  1.238(15)& 0.8078(98)\\
   0.050\ \ &0.774(13) & 1.289(23)&  1.310(22)& 0.761(14) \\
   0.100\ \ &0.748(13) & 1.336(25)&  1.266(22)& 0.789(15) \\
   \hline
   ``chiral limit'':\ \ &0.7325(88)& 1.364(16)&  1.240(15)& 0.8057(97)\\
    ($\chi^2/$dof     &  1.61    &  1.64   &     1.61  &   1.64)   \\
   \hline\hline
  \end{tabular}
 \caption{
   Renormalization factors of the scalar operator and quark mass in
   the RGI and in the $\ovl{\rm MS}$ schemes at $\mu=2$ GeV.
   The results at each sea quark mass are listed as well as those in
   the chiral limit of sea quarks.
   }
 \label{Zfactors}
\end{table}

\begin{table}[tbp]
 \begin{tabular}{rcc}
  \hline\hline
  $m_{\rm sea}(m_{ud})$
  &\ \ \ \ $Z_T^{\rm RGI}$\ \ \ \ 
  & $Z_T^{\ovl{\rm MS}}(2\ {\rm GeV})$\\
  \hline
  NF2$\epsilon$,\  
  0.002\ \ & 1.7023(62)& 1.4689(53)\\
  \hline\hline
  NF2p,\
  0.015\ \ & 1.7461(69)&  1.5066(59)\\
  0.025\ \ & 1.7418(63)&  1.5030(54)\\
  0.035\ \ & 1.7393(70)&  1.5008(61)\\
  0.050\ \ & 1.7470(72)&  1.5075(62)\\
  0.070\ \ & 1.7361(63)&  1.4981(54)\\
  0.100\ \ & 1.7330(64)&  1.4953(55)\\
  \hline
  0.000\ \ & 1.7441(38)& 1.5050(33)\\
  ($\chi^2/$dof&  0.24     &   0.24) \\
  \hline\hline
  NF3p-a,\
  0.015\ \ &  1.7662(44)&  1.5283(38)\\
  0.025\ \ &  1.7663(50)&  1.5284(43)\\
  0.035\ \ &  1.7685(54)&  1.5303(47)\\
  0.050\ \ &  1.7591(38)&  1.5222(33)\\
  0.080\ \ &  1.7674(48)&  1.5294(41)\\
  \hline
  NF3p-b,\
  0.015\ \ &  1.7620(43)&  1.5247(37)\\
  0.025\ \ &  1.7630(42)&  1.5255(37)\\
  0.035\ \ &  1.7637(42)&  1.5261(36)\\
  0.050\ \ &  1.7518(43)&  1.5159(37)\\
  0.100\ \ &  1.7640(56)&  1.5264(48)\\
  \hline
  ``chiral limit'':\ \ &1.7639(35)&  1.5262(30)\\
  ($\chi^2/$dof &  1.18    &   1.18) \\
  \hline\hline
 \end{tabular}
 \caption{Same as Table~\ref{Zfactors} but for $Z_T$.}
 \label{Zfactors_T}
\end{table}

\begin{figure}[tbp]
 \includegraphics[width=8cm,clip]{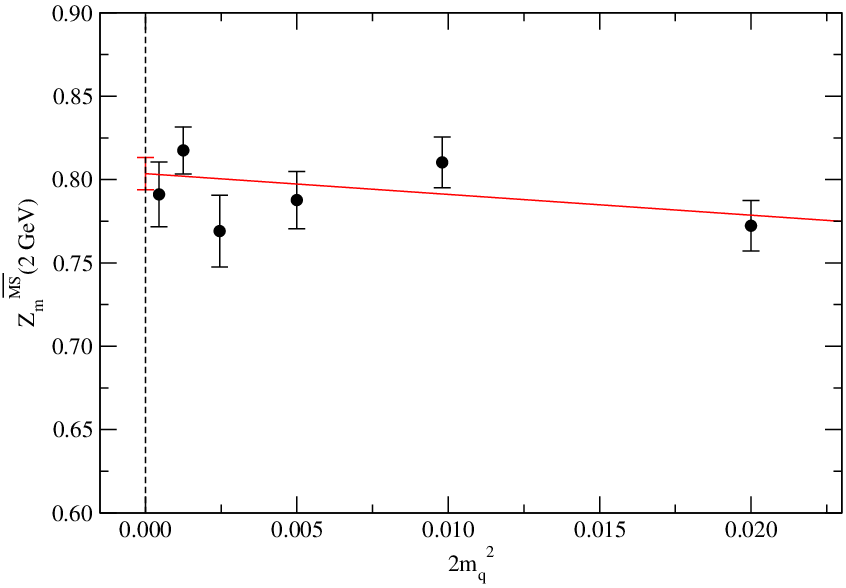}
 \includegraphics[width=8.2cm,clip]{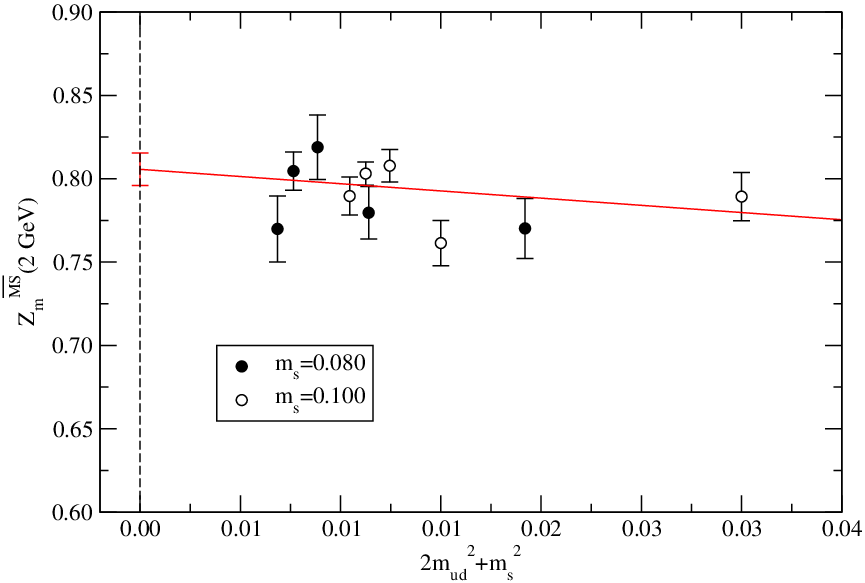}
 \caption{
  Left panel: $Z_m^{\ovl{\rm MS}}(2\ {\rm GeV})$ for NF2p as a function of 
   $2m_{\rm sea}^2$.  
   Right panel: Same value for NF3p-a and NF3p-b as a function of 
   $2m_{ud}^2+m_s^2$. Linear extrapolation to the $m_{\rm sea}=0$ limit or
 the $m_{ud}=0$ limit is shown as well.
 }
 \label{Zmmsb_chi}
\end{figure}

\begin{figure}[tbp]
 \includegraphics[width=8cm,clip]{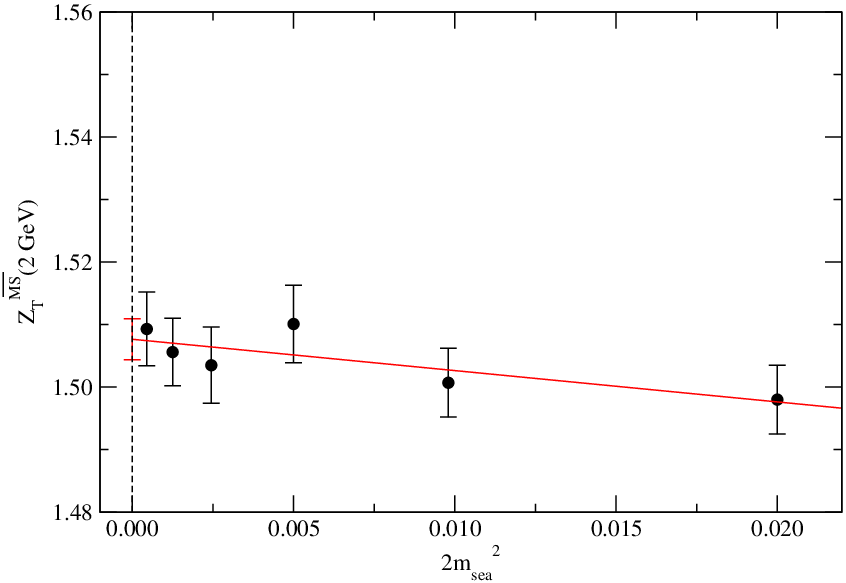}
 \includegraphics[width=8.2cm,clip]{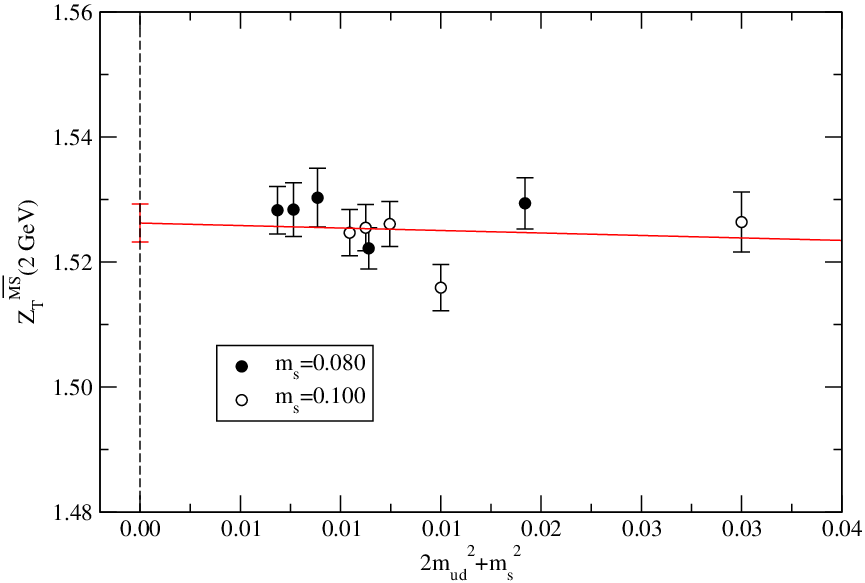}
 \caption{
   Same as Figure~\ref{Zmmsb_chi} but for 
   $Z_T^{\ovl{\rm MS}}(2\ {\rm GeV})$. 
 }
 \label{ZTmsb_chi}
\end{figure}

The renormalization constants in the $\overline{\rm MS}$ scheme are
obtained as
$Z_{\cal O}^{\ovl{\rm MS}}(\mu) = w_{\cal O}^{\ovl{\rm MS}}(\mu)\cdot
 Z_{\cal O}^{\rm RGI}$,
again using the matching factor to the RGI value 
$w_{\cal O}^{\ovl{\rm MS}}(\mu)$.
Results of the RGI value and those in the $\overline{\rm MS}$ scheme at
$\mu$ = 2~GeV are listed in Table~\ref{Zfactors} for ${\cal O}$ = $m$
and $S$ with the four-loop level matching, and in Table~\ref{Zfactors_T} 
for ${\cal O} = T$ with the three-loop level.

For the NF2p ensembles the renormalization factors
at finite sea quark masses are extrapolated to the limit of $m_{\rm sea}
=0$ as a linear function of $2m_{\rm sea}^2$.
For the 2+1-flavor data, we combine NF3p-a and NF3p-b to quote the final
result in the chiral limit of all of the three flavors, assuming a sea quark
mass dependence of the form $Z(1+c^{(3)}(2m_{ud}^2+m_s^2))$.
The extrapolation is shown in Figures~\ref{Zmmsb_chi} and
\ref{ZTmsb_chi} for the NF2p (left panel) and NF3p-a/b (right panel)
ensembles. 
Although we do not observe any systematic sea quark mass dependence,
the data show larger fluctuations than the statistical errors at each
sea quark mass for $Z_m$.
As a result, the $\chi^2/$dof for the combination of NF3p-a and NF3p-b is 
uncomfortably large ($\sim$ 2.6), as listed in Tables~~\ref{Zfactors}.
This may indicate that the statistical error estimated at each sea
quark mass is underestimated.
It is also suggested from the size of the statistical error at a fixed
sea quark mass, say $m_{\rm sea}$ (or $m_{ud}$) = 0.015.
Namely, the size of error is comparable between NF2p and NF3p-a/b,
though the statistics is more than factor of two larger for NF2p.
We use the jackknife method for the statistical analysis with a bin
size of 50 HMC trajectories.
Given the limited total length of trajectories (2,500 for NF3p-a/b), the
statistical error does not change much even if we increase the bin
size to 100 trajectories.
We do not investigate this point further, because the statistical
error does not give the dominant part of the error in the final results.

For the central values of the final results, we quote the result at 
$m_{\rm sea}=0.002$ for NF2$\epsilon$ and that in the $m_{\rm sea} =0$ 
limit for NF2p or in the $m_{ud} =m_s =0$ limit for the combination of 
NF3p-a and NF3p-b.
In Table~\ref{Zfactors}, the extrapolated values are listed in
separated rows. 
The results for $Z_m$ are
\begin{eqnarray}
 Z_m^{\ovl{\rm MS}}(2\ {\rm GeV}) = 
  \left\{
  \begin{array}{l}
   0.824(14)(24)\left({}^{+14}_{-00}\right)\ \ {\rm for\ } N_f=2,\ \beta=2.35\\
   0.804(10)(25)\left({}^{+00}_{-21}\right)\ \ 
    {\rm for\ } N_f=2,\ \beta=2.30\\
   0.806(12)(24)\left({}^{+00}_{-11}\right)\ \ 
    {\rm for\ } N_f=2+1, \beta=2.30,\\
  \end{array}
  \right.\label{final_result}
\end{eqnarray}
The first error is statistical, which includes the small statistical
errors in the extraction of $Z_A^{\rm WTI}$ and the lattice scale $a^{-1}$.
The scale affects the determination of the matching point $\mu=2$ GeV.
On this error, we also take account of the ambiguity in 
removing the scale dependence of $Z_m^{\rm RGI}$ by comparing the
results with different ranges of the linear fit.
The systematic errors given in the second and third parentheses are
described in the following.

An important source of the systematic error is the truncation of the
perturbative expansion in the matching between the RI/MOM and
$\overline{\rm MS}$ schemes.
It is given by a ratio of two matching factors to the Renormalization
Group Invariant (RGI) value, {\it i.e.} $w_m^{\rm RI/MOM}(\mu)$ and
$w_m^{\overline{\rm MS}}(\mu)$ in (\ref{matching_def}).
The perturbative expansion of these factors is given in (\ref{w_factor}) and
known to four-loop order.
By setting $\mu$ = 2~GeV, we may evaluate how it depends on the loop
order. 
For $N_f=2$, 
the ratio $w_m^{\overline{\rm MS}}(\mu)/w_m^{\rm RI/MOM}(\mu)$ becomes
1, 0.911, 0.863, and 0.835 when the perturbative expansion includes
$O(\alpha_s^0)$, $O(\alpha_s^1)$, $O(\alpha_s^2)$, and $O(\alpha_s^3)$
terms respectively.
From this observation, we find that the perturbative expansion
converges such that the additional correction is about 60\% of the
correction of the previous order. Same level of the correction is 
observed for the case of $N_f=2+1$.
We therefore assume that this convergence persists at the next unknown
perturbative coefficient.
The second error in (\ref{final_result}) is estimated by taking 
a difference of
the current best four-loop analysis and the second best three-loop
analysis and multiplying a factor 0.6.

The effect of SCSB may arise in two different ways.
First, in the extraction of $Z_q$ we used the axial-vector
vertex function $\Lambda_A$, but if we used the vector vertex function
$\Lambda_V$ instead the result is slightly shifted, which is given in
the third parentheses.
Note that this does not matter for NF2$\epsilon$, because there is no
significant difference between $\Lambda_V$ and $\Lambda_A$ in the 
$\epsilon$-regime. 
Second, one may expect some uncertainty in the process of subtraction
of the power divergent piece from $\langle\bar{q}q\rangle^{(N)}$.
In Section \ref{PBPtreatments}, we demonstrate that the power-divergent 
term can be removed from $\vev{\bar{q}q}^{(N)}$ to obtain 
$\vev{\bar{q}q}^{\rm (subt)}$ in almost $N$-independent way. 
However, it does not guarantee that the results for 
$\vev{\bar{q}q}^{\rm (subt)}$ are unchanged beyond the maximum value of
$N$ we studied. 
In fact, in the $p$-regime, we find that 
$\Sigma\equiv\vev{\bar{q}q}^{(\rm subt)}(m_q=0)$ obtained with various 
values of $N$ slightly increases as a function of $N$. 
In the calculation based on the chiral perturbation theory  
 on the same ensembles~\cite{Fukaya:2007yv,Fukaya:2009aa},
we find $\approx 10\%$ larger values of $\Sigma$ than those from the eigenvalue 
decomposition for NF2p and NF3p-a/b.
For NF2$\epsilon$, the result of the
calculation based on the chiral random matrix theory~\cite{Fukaya:2007fb}
is $\approx 10\%$ smaller.
To estimate the effect of the truncation of 
eigenvalues, we repeat the same analysis by fixing the value of $\Sigma$
10\% smaller (larger) than the original one for
NF2$\epsilon$ (NF2p and NF3p-a/b).  
As a result, we find the magnitude of finite $N$ effect is similar to 
the statistical errors for all cases. 
We quote the difference from the central value in the third error in 
(\ref{final_result}).
For NF2p and NF3p-a/b, we combine this error with
the effect from the difference between $\Lambda_A$ and $\Lambda_V$,
which is in the same direction.

For completeness, we also present results for $Z_T$. 
Matching procedures are illustrated in Figure~\ref{Matching_T}.
Table~\ref{Zfactors_T} summarizes the RGI and the $\ovl{\rm MS}$ values.
The left and right panels in Figure~\ref{ZTmsb_chi} show the linear 
extrapolation as a function of $2m_{\rm sea}^2$ (NF2p) and 
$2m_{ud}^2 +m_s^2$ (NF3p-a/b), respectively.

\section{Conclusion}~\label{Conclusion}
We calculated the renormalization factors for the quark bilinear
operators constructed from the overlap fermion formulation, based on
the original idea of NPR proposed in~\cite{Martinelli:1994ty}.
The aim of this calculation is to provide the renormalization factors for
a series of numerical studies being performed by the JLQCD and TWQCD
collaborations using dynamical overlap fermions.
By virtue of the exact chiral symmetry of the overlap fermion, the
analysis is largely simplified compared to other non-chiral fermion
formulations. 

Through the simulation in the $\epsilon$-regime, we explicitly confirm
that the vector and axial-vector vertex functions agree with each
other when the effect of spontaneous chiral symmetry breaking is
negligible. 
This may provide a clean way to calculate the renormalization factors
through the NPR method, since the calculation does not suffer from the
potential problems due to pion poles.

In the $p$-regime, where the spontaneous symmetry breaking effectively
remains even on a finite volume lattice, we may precisely control the
non-perturbative quark mass dependence of the quark propagator and
vertex functions using the OPE analysis supplemented by the condensate
explicitly constructed from the low-lying quark eigenmodes.
The exact chiral symmetry of the overlap fermion plays an important
role also in this analysis.

Our main results are those of the mass renormalization factor $Z_m$,
which is an inverse of the scalar density renormalization factor
$Z_S$.
The result has been already used in the calculation of the chiral condensate
in two-flavor QCD from the Dirac operator 
spectrum~\cite{Fukaya:2007fb,Fukaya:2007pn} and from
the topological susceptibility~\cite{Chiu:2008jq}.
It has also been used in our calculation of up and down quark mass through
the analysis of pion mass and decay constant~\cite{Noaki:2008iy}.
Extension of these works to the 2+1-flavor case is in progress.

By using the value of $Z_m^{\ovl{\rm MS}}$ we quoted in this article,
we are planning to determine the up and down quark mass $m_{ud}$ and the 
strange quark mass $m_s$ from 
the analysis of the meson masses $m_\pi^2$ and $m_K^2$ and the decay 
constants $f_\pi$ and $f_K$ in the $N_f=2+1$ dynamical simulation. 
A preliminary results from this project was reported in~\cite{Noaki:2008gx}.

\begin{acknowledgments}
Numerical simulations are performed on Hitachi SR11000 and
IBM System Blue Gene Solution at High Energy Accelerator Research
Organization (KEK) under a support of its Large Scale
Simulation Program (Nos.~07-16 and~08-05 ). We thank Prof. C. Sachrajda for 
informative discussions.
HF is supported in part by the Global COE Program 
``Quest for Fundamental Principles in the Universe'' of Nagoya University 
provided by Japan Society for the Promotion of Science (G07).
This work is supported in part by the Grant-in-Aid of the 
Ministry of Education
(Nos.
18740167,
18840045,
19540286,
19740121,
19740160,
20025010,
20039005,
20740156,
20105002,
20105005,
21674002)
and the National Science Council of Taiwan (No. NSC96-2112-M-002-020-MY3)
and NTU-CQSE (Nos. 97R0066-65 and 97R0066-69).

\end{acknowledgments}

\appendix

\section{Perturbative matching}
\label{matching}

In this appendix, we present the details of our matching procedure.

The matching of an operator ${\cal O}$ between the $\overline{\rm MS}$
scheme and the RI/MOM scheme is written as
\begin{equation}
  {\cal O}^{\overline{\rm MS}}(\mu) = 
  \frac{
    w_{\cal O}^{\ovl{\rm MS}}(\mu)
  }{
    w_{\cal O}^{\rm RI/MOM}(\mu_0)
  }
  {\cal O}^{\rm RI/MOM}(\mu_0),\label{matching_def}
\end{equation}
where the conversion factor $w_{\cal O}^{X}(\mu)$ from a given scheme $X$
to the so-called Renormalization Group Invariant (RGI) value is
written as
\begin{eqnarray}
 w_{\cal O}^X(\mu) &=&
  \alpha_S(\mu)^{\bar{\gamma}_0}
  \Biggl[
  1 +(\bar{\gamma}_1 -\bar{\beta}_1\bar{\gamma}_0)
  \frac{\alpha_S(\mu)}{4\pi}\nn\\
 & &
  +\tfrac{1}{2}
  \left(
   (\bar{\gamma}_1 -\bar{\beta}_1\bar{\gamma}_0)^2
   +\bar{\gamma}_2 +\bar{\beta}_1^2\bar{\gamma}_0
   -\bar{\beta}_1\bar{\gamma}_1 -\bar{\beta}_2\bar{\gamma}_0
  \right)
  \left(\frac{\alpha_S(\mu)}{4\pi}\right)^2\nn\\
 & &
  +\Bigl(
    \tfrac{1}{6}(\bar{\gamma}_1 -\bar{\beta}_1\bar{\gamma}_0)^3
    +\tfrac{1}{2}(\bar{\gamma}_1 -\bar{\beta}_1\bar{\gamma}_0)
    (\bar{\gamma}_2 +\bar{\beta}_1^2\bar{\gamma}_0 
    -\bar{\beta}_1\bar{\gamma}_1 -\bar{\beta}_2\bar{\gamma}_0)\nn\\
 & &
  +\tfrac{1}{3}(
  \bar{\gamma}_3 -\bar{\beta}_1^3\bar{\gamma}_0 
  +2\bar{\beta}_1\bar{\beta}_2\bar{\gamma}_0 -\bar{\beta}_3\bar{\gamma}_0
  +\bar{\beta}_1^2\bar{\gamma}_1 -\bar{\beta}_2\bar{\gamma}_1
  -\bar{\beta}_1\bar{\gamma}_2)
   \Bigr)
  \left(\frac{\alpha_S(\mu)}{4\pi}\right)^3
  \Biggr]_X,\label{w_factor}
\end{eqnarray}
to the four-loop order in terms of the strong coupling constant
$\alpha_S(\mu)$.
(For the coupling constant, we always use the $\overline{\rm MS}$ scheme.)
The coefficients $\bar{\beta}_i$ and $\bar{\gamma}_i$ are given in
terms of the coefficients of the $\beta$-function $\beta(\alpha_S)$
and the anomalous dimension $\gamma_{\cal O}(\alpha_S)$
\begin{eqnarray}
 \beta &=& -\beta_0\frac{\alpha_S^2}{4\pi}
         -\beta_1\frac{\alpha_S^3}{(4\pi)^2}
         -\beta_2\frac{\alpha_S^4}{(4\pi)^3}
         -\beta_3\frac{\alpha_S^5}{(4\pi)^4}-\cdots,\\
 \gamma_{\cal O} &=& -\gamma_{\cal O}^{(0)}\frac{\alpha_S}{4\pi}
                   -\gamma_{\cal O}^{(1)}\left(\frac{\alpha_S}{4\pi}\right)^2
                   -\gamma_{\cal O}^{(2)}\left(\frac{\alpha_S}{4\pi}\right)^3
                   -\gamma_{\cal O}^{(3)}\left(\frac{\alpha_S}{4\pi}\right)^4
		   -\cdots.
\end{eqnarray}
as $\bar{\beta}_i=\beta_i/\beta_0$ and 
$\bar{\gamma}_i=\gamma_{\cal O}^{(i)}/\beta_0$.

The $\beta$-function is specified by
\begin{eqnarray}
 \beta_0 &=& 11-\frac{2}{3}N_f,\\
 \beta_1 &=& 102-\frac{38}{3}N_f,\\
 \beta_2 &=& \frac{2857}{2}-\frac{5033}{18}N_f+\frac{325}{54}N_f^2,\\
 \beta_3 &=& \frac{149753}{6} +3564\zeta_3
 -\left(\frac{1078361}{162} +\frac{6508}{27}\zeta_3 \right)N_f\nn\\
 & &
 +\left(\frac{50065}{162} +\frac{6472}{81}\zeta_3 \right)N_f^2
 +\frac{1093}{729}N_f^3,
\end{eqnarray}
with $\zeta_3$ = 1.2020569.
The running coupling constant is then obtained as 
\begin{eqnarray}
 \alpha_S(\mu) &=&
  \frac{4\pi}{\beta_0\ln\mu^2/\Lambda^2}
  \Biggl[
  1-\frac{\beta_1}{\beta_0^2}
  \frac{\ln\left(\ln\mu^2/\Lambda^{2}\right)}
  {\ln\mu^2/\Lambda^{2}}\nn\\
 & &
  +\frac{\beta_1^2}{\beta_0^4\left(\ln\mu^2/\Lambda^{2}\right)^2}
  \left(
   \left(\ln\ln\mu^2/\Lambda^{2}\right)^2
   -\ln\left(\ln\mu^2/\Lambda^{2}\right)
   +\frac{\beta_2\beta_0}{\beta_1^2} -1
  \right)\nn\\
& &
 -\frac{\beta_1^3}{\beta_0^6\left(\ln\mu^2/\Lambda^{2}\right)^3}
 \Biggl(
  \left(\ln\ln\mu^2/\Lambda^{2}\right)^3
  -\frac{5}{2}\left(\ln\ln\mu^2/\Lambda^{2}\right)^2\nn\\
& &
 \hspace{3.9cm}
  -\left(2 -\frac{3\beta_0\beta_2}{\beta_1^2}\right)
                    \ln\ln\mu^2/\Lambda^{2}
	       +\frac{1}{2} -\frac{\beta_0^2\beta_3}{2\beta_1^3}\Biggr)
    \Biggr].
\end{eqnarray}
In our work, we chose $\Lambda$ = 245~MeV for both 2 and 2+1-flavor
analysis. 

The renormalization of the scalar bilinear operator 
${\cal O}=S=\bar{q}q$ is an inverse of the mass renormalization.
The anomalous dimension thus has a relation 
$\gamma_m = -\gamma_{\rm S}$.  
At the lowest order, $\gamma_m^{(0)} =4$ for any scheme.
Higher order coefficients are calculated in \cite{Chetyrkin:1999pq}
to the four-loop order in the RI/MOM scheme
\begin{eqnarray}
 \gamma_m^{(1)} &=& 126 -\frac{52}{9}N_f,\\
 \gamma_m^{(2)} &=& \frac{20911}{3} -\frac{3344}{3}\zeta_3
  +\left(-\frac{18386}{27} +\frac{128}{9}\zeta_3\right)N_f
  +\frac{928}{81}N_f^2,\\
 \gamma_m^{(3)} &=& \frac{300665987}{648} 
  -\frac{15000871}{108}\zeta_3 +\frac{6160}{3}\zeta_5
  +\left(
    -\frac{7535473}{108} +\frac{627127}{54}\zeta_3 +\frac{4160}{3}\zeta_5
   \right)N_f\nn\\
 & &
  +\left(
    \frac{670948}{243} -\frac{6416}{27}\zeta_3
   \right)N_f^2 
  -\frac{18832}{729}N_f^3,
\end{eqnarray}
and in the $\ovl{\rm MS}$ scheme
\begin{eqnarray}
 \gamma_m^{(1)} &=& \frac{202}{3} -\frac{20}{9}N_f,\\
 \gamma_m^{(2)} &=& 1249 
  -\left(\frac{2216}{27} +\frac{160}{3}\zeta_3\right)N_f -\frac{140}{81}N_f^2,\\
 \gamma_m^{(3)} &=& \frac{4603055}{162} +\frac{135680}{27}\zeta_3 -8800\zeta_5 
  -\left(\frac{91723}{27} +\frac{34192}{9}\zeta_3 -880\zeta_4
    -\frac{18400}{9}\zeta_5
   \right)N_f\nn\\
 & &
  +\left(
    \frac{5242}{243} +\frac{800}{9}\zeta_3 -\frac{160}{3}\zeta_4 
   \right)N_f^2
  -\left(\frac{332}{243} -\frac{64}{27}\zeta_3\right)N_f^3,
\end{eqnarray}
where
$\zeta_3 = 1.202057$, $\zeta_4 = \pi^4/90$ and $\zeta_5 = 1.036928$.

For the quark field renormalization (${\cal O}=q$), the lowest order
coefficient vanishes in the Landau gauge for any scheme. 
The higher order coefficients for the RI/MOM scheme are
\cite{Chetyrkin:1999pq}
\begin{eqnarray}
 \gamma_q^{(1)} &=& \frac{67}{3} -\frac{4}{3}N_f,\\
 \gamma_q^{(2)} &=& \frac{43477}{36} -\frac{607}{2}\zeta_3
  -\left(\frac{3674}{27} -16\zeta_3 \right)N_f +\frac{80}{27}N_f^2,\\
 \gamma_q^{(3)} &=& \frac{54714743}{648} -\frac{7004309}{162}\zeta_3
  +\frac{15846715}{1296}\zeta_5
 -\left(\frac{4659455}{324} -\frac{637413}{162}\zeta_3 +830\zeta_5\right)N_f
 \nn\\
 & &
 +\left(\frac{166269}{243} -64\zeta_3\right)N_f^2 -\frac{688}{81}N_f^3,
\end{eqnarray}
while the $\ovl{\rm MS}$ coefficients are 
\begin{eqnarray}
 \gamma_q^{(1)} &=& \frac{67}{3} -\frac{4}{3}N_f,\\
 \gamma_q^{(2)} &=& \frac{20729}{36} -\frac{79}{2}\zeta_3
 -\frac{550}{9}N_f +\frac{20}{27}N_f^2,\\
 \gamma_q^{(3)} &=& \frac{2109389}{162} -\frac{565939}{324}\zeta_3
  +\frac{2607}{4}\zeta_4 -\frac{761525}{1296}\zeta_5
  -\left(\frac{162103}{81} +\frac{2291}{27}\zeta_3 +\frac{79}{2}\zeta_4 
    +\frac{160}{3}\zeta_5 \right)N_f\nn\\
 & &
  +\left(\frac{3853}{81} +\frac{160}{9}\zeta_3\right)N_f^2
  +\frac{140}{243}N_f^3.
\end{eqnarray}

Coefficients for the tensor current ${\cal O}=T=\bar{q}\sigma_{\mu\nu}q$
are known to three-loop \cite{Gracey:2003yr,Aoki:2007xm}.
Besides the common value $\gamma_T^{(0)} = 4/3$, higher order coefficients 
for the RI/MOM scheme are
\begin{eqnarray}
 \gamma_T^{(1)} &=& \frac{362}{9} -\frac{52}{27}N_f,\\
 \gamma_T^{(2)} &=& \frac{159607}{81} -\frac{13072}{27}\zeta_3
  +\left(-\frac{17426}{81} +\frac{256}{27}\zeta_3 \right)N_f 
  +\frac{928}{243}N_f^2,
\end{eqnarray}
and those for the $\ovl{\rm MS}$ scheme are 
\begin{eqnarray}
 \gamma_T^{(1)} &=& \frac{362}{9} -\frac{52}{27}N_f,\\
 \gamma_T^{(2)} &=& \frac{52555}{81} -\frac{928}{27}\zeta_3
  -\left(\frac{5240}{81} +\frac{160}{9}\zeta_3 \right)N_f 
  -\frac{4}{9}N_f^2.
\end{eqnarray}

\bibliography{NPRref}

\end{document}